\documentclass{emulateapj}
\usepackage{natbib}
\usepackage{threeparttable}

\shorttitle{Connections between star cluster populations and their host galaxy nuclear rings}

\shortauthors{Chao Ma, Richard de Grijs, and Luis C. Ho}

\slugcomment{Accepted for publication in The Astrophysical Journal}

\begin{document}

\title{Connections between star cluster populations and their host galaxy nuclear rings}

\author{Chao Ma\altaffilmark{1,2}, Richard de Grijs\altaffilmark{2,1,3,4},
  and Luis C. Ho\altaffilmark{2,1}}
\altaffiltext{1}{Department of Astronomy, Peking University, Yi He
  Yuan Lu 5, Hai Dian District, Beijing 100871, China}
\altaffiltext{2}{Kavli Institute for Astronomy and Astrophysics,
  Peking University, Yi He Yuan Lu 5, Hai Dian District, Beijing
  100871, China; machao@pku.edu.cn, grijs@pku.edu.cn}
\altaffiltext{3}{International Space Science Institute--Beijing, 1
  Nanertiao, Zhongguancun, Hai Dian District, Beijing 100190, China}
\altaffiltext{4}{Present address: Department of Physics and Astronomy,
  Macquarie University, Balaclava Road, North Ryde, NSW 2109,
  Australia}

\begin{abstract}
Nuclear rings are excellent laboratories for probing diverse phenomena
such as the formation and evolution of young massive star clusters
(YMCs), nuclear starbursts, as well as the secular evolution and
dynamics of their host galaxies. We have compiled a sample of 17
galaxies with nuclear rings, which are well resolved by
high-resolution {\sl Hubble} and {\sl Spitzer Space Telescope}
imaging. For each nuclear ring, we identified the ring star cluster
population, along with their physical properties (ages, masses,
extinction values). We also determined the integrated ring properties,
including the average age, total stellar mass, and current
star-formation rate (SFR). We find that Sb-type galaxies tend to have
the highest ring stellar mass fraction with respect to the host
galaxy, and this parameter is correlated with the ring's SFR surface
density. The ring SFRs are correlated with their stellar masses, which
is reminiscent of the main sequence of star-forming galaxies. There
are striking correlations between star-forming properties (i.e., SFR
and SFR surface density) and non-axisymmetric bar parameters,
appearing to confirm previous inferences that strongly barred galaxies
tend to have lower ring SFRs, although the ring star-formation
  histories turn out to be significantly more complicated. Nuclear
rings with higher stellar masses tend to be associated with lower
cluster mass fractions, but there is no such relation with the ages of
the rings. The two youngest nuclear rings in our sample, NGC 1512
  and NGC 4314, which have the most extreme physical properties,
  represent the young extremity of the nuclear ring age distribution.
\end{abstract}

\section{INTRODUCTION}

In the local Universe, the observed bar fraction is around two-thirds
among normal, luminous galaxies
\citep[e.g.,][]{mulchaey1997,knapen1999,eskridge2000,laurikainen2004,marinova2007}.
Stellar bars, which can redistribute disk material via
non-axisymmetric gravitational torques, are one of the essential
internal drivers of the secular evolution of disk galaxies \citep[for
  a review, see e.g.][and references therein]{kormendy2004}. Such
bar-driven secular processes can cause the inflow of gas from the
outer disk to the central regions, possibly trigging nuclear
starbursts \citep[e.g.][]{sheth2005}. This usually leads to the
formation of interesting substructures in the central regions of the
galaxies, including circumnuclear starburst rings
\citep[e.g.][]{mazzuca2008} as well as nuclear spirals
\citep[e.g.][]{shlosman1990,van2010} inside these rings, which may
represent a channel for gas infall to fuel the central supermassive
black hole (SMBH).

The great majority of nuclear rings are believed to form as a result
of gas inflow along shocks delineated by dust lanes at the leading
edge of the bar \citep[e.g.][]{athan1992}. This gaseous material will
be driven to the circumnuclear region of the host galaxy after angular
momentum loss \citep[e.g.][]{combes1985} by the bar's non-axisymmetric
gravitational potential. The inflowing material would accumulate
around those radii at which the stellar orbits experience dynamical
resonances with the rotating bar potential
\citep[e.g.][]{binney2008}. In the inner galactic region, it is
typically located at so-called inner Lindblad resonances (ILRs) which
arise from the interplay between the bar and the stellar orbits
\citep[e.g.][]{buta1996,boker2008}. As the inflowing gas migrates
toward the ILRs, it gradually speeds up in the azimuthal direction,
and the trajectory gradually changes from radial to nearly circular
\citep[e.g.][]{shlosman2001}. This allows for the accumulation of gas,
yielding ring-like structure surrounding the nucleus.

When the molecular gas density exceeds a certain critical value,
massive star-forming activity would be initiated to form the nuclear
rings we see today. In some unbarred galaxies, other large-scale
non-axisymmetric features such as ovals and strong spiral arms, have
similar dynamical effects as bars in producing nuclear rings
\citep[e.g.][]{combes2001,kormendy2004,mazzuca2006,laurikainen2009}.
Meanwhile, some dusty red nuclear rings have been identified in
elliptical and early-type lenticular galaxies exhibiting little or no
star formation \citep[e.g.][]{wozniak1995}. They may be generated by
the inside-out depletion of dusty nuclear disks
\citep[e.g.][]{lauer2005,comeron2010}.

Nuclear rings are among the most intense star-forming regions in
normal disk galaxies. They often dominate the entire star-formation
activity of their host galaxies \citep[e.g.][]{mazzuca2008}. They are
primarily found in barred Sa--Sc-type spiral galaxies. Characterized
by intense starbursts and by responding to large-scale dynamics,
nuclear rings can dramatically alter the structure of their host
galaxies \citep[e.g.][]{kormendy2004}. Therefore, they are commonly
used as a tracer of the recent inflow of material and they affect the
appearance of pseudobulges \citep[e.g.][]{knapen2006}. In addition,
the properties of nuclear rings can be used as indicators to constrain
galaxy-wide parameters \citep[e.g.][]{weiner2001,li2015}. High
resolution, multi-passband {\sl Hubble Space Telescope} ({\sl HST})
observations of nearby galaxies have shown that the starburst
environments in nuclear rings can harbor large populations of young
massive star clusters (YMCs)
\citep[e.g.,][]{barth1995,buta2000,maoz2001,grijs2012,grijs2017}.
Their ages and masses can be determined based on a careful analysis of
their integrated spectral energy distributions (SEDs)
\citep[e.g.][]{tinsley1968,leitherer1999,anders2004a}. Such star
clusters, and in particular the YMCs, are therefore crucial in probing
a ring's dynamical and star-formation properties
\citep[e.g.,][]{maoz1996,maoz2001,grijs2012,van2013,vaisa2014}.

In this paper, we have performed an extensive statistical analysis of
a sample of nuclear rings, to understand how ring
environments/properties affect the early evolution of their cluster
populations and to explore whether the derived integrated ring
properties---such as the star-formation rate (SFR), the star-formation
density, and the total stellar mass---are linked to any galaxy-wide
parameters (including Hubble type, stellar mass, etc.). We have
compiled a statistically carefully selected sample of nuclear rings
which have already been observed and are well-resolved by both the
{\sl HST} and the {\sl Spitzer Space Telescope}. For each ring in our
catalog, {\sl HST} multi-passband imaging covers the optical
wavelength range in at least four filters, so as to derive reliable
physical cluster parameters \citep[e.g.][]{anders2004a}. As
demonstrated by \citet[][hereafter Paper I]{ma2017}, our newly
developed approach to determining integrated nuclear ring properties
represents a significant improvement in our measurement accuracy
compared with previous efforts. We can thus determine numerous
physical properties of our nuclear rings, including their SEDs, SFRs,
and the average age and total stellar masses. We will combine these
results with the properties of the star cluster populations to
investigate any relevant physical correlations.

This paper has the following structure. In Section 2, we present the
information about both our sample selection and the data
reduction. Section 3 focuses on the technical methods used for
determining the physical parameters of both the rings and the
associated star clusters. Section 4 is devoted to a discussion of how
the derived ring parameters are related to the properties of the ring
cluster populations and their host galaxies. We summarize our results
and conclusions in Section 5.

\section{Ring sample selection and data reduction}
\subsection{Sample selection}

In order to obtain a statistically significant nuclear ring sample, we
extensively mined the literature for suitable target galaxies. Most of
our final sample is based on the compilation of
\citet{comeron2010}. From our initial sample of 108 galaxies
exhibiting nuclear rings, we selected those that had been observed and
resolved by any {\sl HST} camera, including the Wide Field Camera 3
(WFC3), the Wide Field and Planetary Camera 2 (WFPC2), the Advanced
Camera for Surveys (ACS), the Near-Infrared Camera and Multi-Object
Spectrometer (NICMOS), and the Faint Object Camera (FOC). We then
collected their broad- and medium-band images from the {\sl HST}
Legacy Archive (HLA)\footnote{http://hla.stsci.edu/hlaview.html} in as
many filters as possible, ensuring that they covered the largest
available wavelength range. Our multi-wavelength imaging data set was
pipeline-processed and calibrated using the standard HLA reduction
software.

We excluded rings from our selection based on the following criteria:
(i) inclination angle $i>70\arcdeg$; (ii) central regions
  exhibiting dusty features which are classified as star-forming
  rings, which may require infrared or radio observations to provide
  direct access to the star formation activity. Since very few
  clusters are detected in {\sl HST} optical imaging of such target
  galaxies (particularly in the $U$ band), these configurations
  prevent us from studying the cluster population; (iii) dusty red
  rings with no signs of star formation, usually identified in
  elliptical galaxies; the formation of such rings may have nothing to
  do with dynamical resonances but is probably related to the
  inside-out depletion of dusty nuclear disks
  \citep[][]{lauer2005}. Note that such rings should be distinguished
  from those described by criterion (ii) because of the different
  formation mechanisms involved; (iv) no well-defined ring morphology
  in optical passbands: \citet[][]{comeron2010} identified nuclear
  rings through visual inspection of archival {\sl HST} imaging, which
  is not equivalent to a quantitative approach to sample
  selection. Therefore, the radial profiles of some rings (i.e., the
  isophotal intensity as a function of the radial range covered by the
  ring) are too shallow to be detected and distinguished from adjacent
  nuclear discs in broad {\sl HST} passbands, thus yielding
  unreasonable {\sc galfit} 2D component fits. We rejected such
  rings. 

Figure 1 includes examples of composite images of three central rings
that were excluded from our original sample based on these selection
criteria. The left-hand panel shows the red dusty nuclear ring in the
elliptical galaxy NGC 3258. Since only two {\sl HST} optical filters
are available for this galaxy, the F435W and F814W images are rendered
in blue and red, respectively. The ring in the middle panel, located
in strongly barred galaxy NGC 1300, does not show a well-defined ring
structure based on its radial profile, while the right-hand panel
presents the ring of NGC 4459, which is obscured by dust.

\begin{figure}
\begin{center}
\includegraphics[width=1.0\columnwidth]{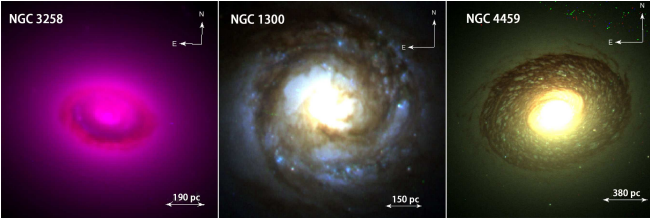}
\caption{ Examples of rejected rings:  (Left) Red dusty ring in the elliptical galaxy NGC 3258;
  the F435W and F814W images are rendered in blue and red,
  respectively. (Middle) Nuclear ring of NGC 1300, which does not show
  a well-defined ring morphology. Its color composite image was
  obtained by stacking individual images in the F435W, F555W, and
  F814W filters. (Right) The ring in NGC 4459 is highly obscured by
  dust. The galaxy's F435W, F555W, and F814W images are rendered in
  blue, green, and red, respectively.}
\end{center}
\end{figure}

The final sample consists of 17 galaxies, for which we have included
the relevant information in Table 1.  Figure 2 shows the composite
images of our sample rings. The $U$, $V$, and $I$ images are shown in
blue, green, and red, respectively. The $U$ band (either F330W or
F336W, depending on the {\sl HST} camera used) primarily traces the
young stellar populations; nuclear rings are most prominent in
$U$-band filters. Images observed through the corresponding
narrow-band H$\alpha$ filters were also retrieved if available. From
the {\sl Spitzer} Heritage Archive we downloaded any available
broad-band infrared images (post-basic calibrated data, with a pixel
scale of 0.6\arcsec) observed with the {\sl Spitzer}/Infrared Array
Camera (IRAC). Table 1 includes the relevant observational details and
host galaxy's parameters.

Most of our sample galaxies are early-type (S0$^{+}$--Sbc) barred
spirals, reaffirming previous studies that nuclear star formation is
preferentially observed in early-type systems. \citet[][]{ho1997}
argued that the enhancement of star formation in the centers of
early-type barred spirals can be explained in terms of the structural
differences between bars in early- and late-type spirals. Three sample
galaxies---NGC 7217, NGC 7742, and UGC 3789---are known to be
unbarred, which thus casts doubt on the notion that a bar may be an
essential driver to form the nuclear ring. For both NGC 7217 and NGC
7742, both nuclear rings and counter-rotating components (with respect
to the host stellar background disk) have been reported (see de Zeeuw
et al. 2002 and Merrifield \& Kuijken 1994 for NGC 7742 and NGC 7217,
respectively). This can be interpreted as having resulted from a past
minor merger with a gas-rich dwarf galaxy on a retrograde orbit
\citep[][]{silchenko2006}, and the resulting merger-induced ovals
could have had similar dynamical effects as those associated with
bars, thus potentially leading to ring formation. In the presence of
these counter-rotating components, the corresponding ring morphologies
tend to be relatively more circular than those in barred galaxies, as
shown in Figure 2.

The final column of Table 1 lists the non-axisymmetric torque
parameter, $Q_{\rm g}$, which quantifies the strength of
non-axisymmetric perturbations in a galaxy
\citep[][]{combes1981,buta2001}. It is defined as the highest value of
the tangential forces normalized by the axisymmetric force field and
is expressed as
\begin{equation}
Q_{\rm g}={\rm max}( \frac{|F_{\rm T}(r)|_{\rm max}}{|F_{\rm R}(r)|}),
\end{equation}
where the numerator $|F_{\rm T}(r)|_{\rm max}$ is the maximum
tangential force at a given radius and $|F_{\rm R}(r)|$ symbolizes the
average radial force at the same radius. The $Q_{\rm g}$ values
provide a clean measurement of the total non-axisymmetric bar
strength, with higher values associated with more significant
bars. For galaxies with a strong spiral pattern but a weak or no bar,
this parameter mainly reflects the spiral strength. Lower values of
$Q_{\rm g}$ are statistically related to spiral-arm perturbations and
oval distortions.

The $Q_{\rm g}$ values in this paper were taken from
\citet[][]{comeron2010}, who performed extensive and consistent
measurements for their much larger ring sample based on $H$-band
images from the Two Micron All Sky Survey (2MASS). The $Q_{\rm g}$
value for UGC 3789 has not been reported. Although our sample is
ultimately limited by available number of galaxies, the $Q_{\rm g}$
values in our sample cover a large range, from 0.026 to 0.432, in
broad agreement with the range covered by the sample of
\citet[][]{comeron2010}. It appears, therefore, that our sample is a
diverse and representative subsample of the overall population of
nuclear-ring galaxies.

\begin{figure*}[ht!]
\begin{center}
\includegraphics[width=1.7\columnwidth]{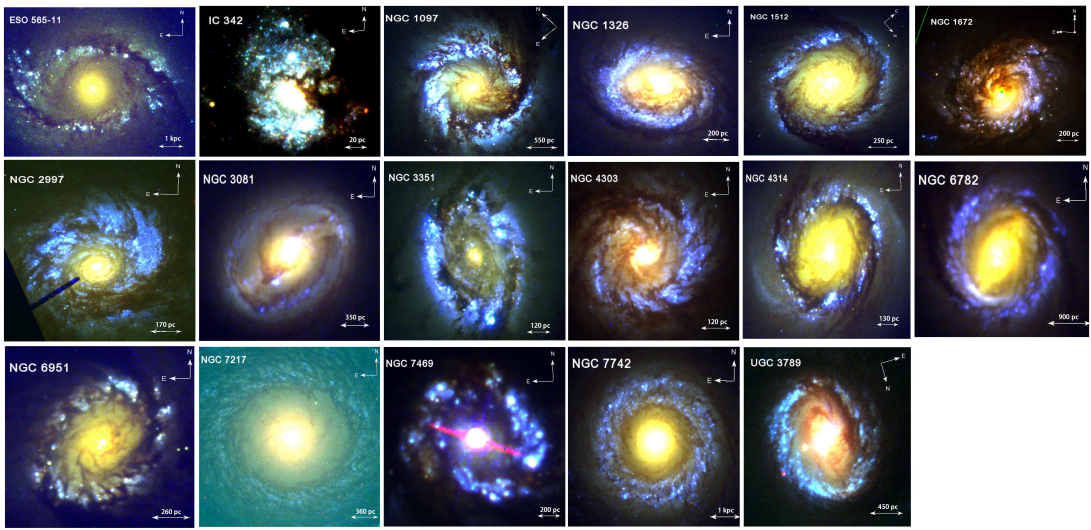}
\caption{{\sl HST} color composite images of our final nuclear ring
  sample, with $U$, $V$, and $I$ images represented in blue, green,
  and red, respectively.}
\label{}
\end{center}
\end{figure*}

\begin{table*}
\begin{threeparttable}
\begin{center}
\begin{minipage}{160mm}
\caption{Observed filter sets and basic galactic parameters of our
  target galaxies}
\label{comparison.tab}
\begin{tabular}{lcclc}
\hline
\hline \\[-1.4ex]
\colhead{Galaxy}&\colhead{Hubble Type}&\colhead{Distance Modulus (mag) }&\colhead{Filters}&$Q_{\rm g}$\\[0.8ex]
\hline\\[-1.4ex]

ESO 565-11&(R)SB(r)0/a&34.23&F255W, F336W, F439W, F555W, F814W&0.316 \\
 IC 342&SAB(rs)cd&27.54&F275W, F336W, F438W, F547M, F814W&0.177  \\
NGC 1097&SB(s)b&31.40&F336W, F438W, F547M, F658N, F814W, 8 $\mu$m&0.241 \\
NGC 1326&(R)SB$0^{+}$(r)&30.86&F255W, F336W, F439W, F555W, F814W&0.163  \\
NGC 1512&SB(r)a&30.48& F336W, F438W, F555W, F658N, F814W, 3.6 $\mu$m, 8 $\mu$m&0.366 \\
NGC 1672&SB(s)b&30.81&F330W, F435W, F550M, F606W,  F814W&0.349 \\
NGC 2997&SAB(rs)c&30.2&F220W, F330W, F336W, F555W, F606W, F814W, 8 $\mu$m&0.306 \\
NGC 3081&(R)SAB(r)0/a&32.09&F255W, F336W, F439W, F555W, F814W&0.194  \\
NGC 3351&SB(r)b&30.0&F275W, F336W, F438W, F450W, F555W, F606W, F657N &0.311   \\
&&&F814W, 3.6 $\mu$m, 4.5 $\mu$m, 5.8 $\mu$m, 8 $\mu$m\\
NGC 4303&SAB(rs)bc&30.91&F218W, F330W, F555W, F814W, 8 $\mu$m&0.285\\
NGC 4314&SB(rs)a&29.93&F336W, F439W, F569W, F606W, F658N, F814W, 8 $\mu$m&0.432   \\
NGC 6782&(R)SAB(r)a&33.61&F255W, F300W, F450W, F606W, F814W&0.205\\
NGC 6951&SAB(rs)bc&31.77&F110W, F160W, F330W, F547M, F606W, F658N, F814W&0.275\\&&& 8 $\mu$m \\
NGC 7217&(R)SA(r)ab&31.41&F336W, F450W, F547M, F606W, F658N, F814W, 8 $\mu$m&0.026 \\
NGC 7469&(R')SAB(rs)a&34.07&F330W, F435W, F550M, F606W, F814W&0.049 \\
NGC 7742&SA(r)b&32.91&F336W, F555W, F675W, F814W, 5.8 $\mu$m, 8 $\mu$m&0.055 \\
UGC 3789&(R)SA(r)ab&33.49&F110W, F160W, F336W, F438W, F814W&------  \\
\hline

\end{tabular}
\begin{tablenotes}
\item {\bf Notes}. The distance moduli represent the geometric means
  of 10 individual distance measurements compiled in the NASA
  Extragalactic Database (NED; http://ned.ipac.caltech.edu). The
  morphological types were also obtained from NED, while the values of
  the non-axisymmetric bar parameter $Q_{\rm g}$ were collected from
  \citet[][]{comeron2010}.
\end{tablenotes}
\end{minipage}
\end{center}
\end{threeparttable}
\end{table*}

\subsection{Data reduction}

We determine the current (i.e., more recent than 10 Myr) SFR in any
nuclear ring using the calibration recipe of
\citet[][]{kennicutt2009},
\begin{equation}
\rm{SFR}\, (M_{\odot}\, yr^{-1}) = 5.5 \times 10^{-42} [{\it L}
  (H\alpha)_{\rm{obs}} + 0.011{\it L}(8\,{\mu}m)],
\end{equation}
where $L$(H$\alpha$)$_{\rm obs}$ and $L(8\,\mu$m) are the observed
H$\alpha$ (uncorrected for internal dust attenuation) and {\sl
  Spitzer} 8 $\mu$m polycyclic aromatic hydrocarbon (PAH)
emission-line luminosities, respectively, both in units of erg
s$^{-1}$. These luminosities can be derived from the
continuum-subtracted {\sl HST}/H$\alpha$ (e.g., F658N) narrow-band
filters and {\sl Spitzer} 8 $\mu$m bands (for technical details, see
Paper I). H$\alpha$ emission traces the presence of young massive
stars and is traditionally used as current SFR indicator for
star-formation timescales of $\sim$3--10 Myr
\citep[e.g.][]{hao2011}. However, the downside of using the H$\alpha$
flux alone as a tracer of star-forming regions is that H$\alpha$
emission is very sensitive to dust extinction, and the missing
H$\alpha$ flux that has been absorbed and scattered by dust would be
re-emitted at infrared wavelengths. Therefore, combination of the
H$\alpha$ and 8 $\mu$m luminosities offers a better choice for
calculating the SFR. Six of our sample galaxies (NGC 1512, NGC 1097,
NGC 3351, NGC 4314, NGC 6951, and NGC 7217) were observed through both
H$\alpha$ and infrared (3.6 $\mu$m and 8 $\mu$m) filters, so that we
could determine their ring SFRs.

\section{Method and analysis}

Before understanding how the ring environment influences the early
evolution of the cluster population, we need to characterize the
physical properties of the star clusters and their host galaxies'
nuclear rings. In this section, we describe the methods used for
collecting the star cluster catalogs and performing their photometry,
and we summarize the technique used to estimate the cluster masses and
ages. We then briefly present our recently improved approach to
deriving the integrated ring properties (see also Paper I for details)
and discuss the results obtained at the end of this section.

\subsection{Cluster detection and photometry}

We used customized {\sc idl}\footnote{The Interactive Data Language
  ({\sc idl}) is licensed by Research Systems Inc., of Boulder, CO,
  USA.} procedures on two adjacent {\sl HST} filters occupying the
middle wavelength range of the available filter coverage to find the
ring cluster candidates
\citep[e.g.][]{deGrijs2013a,li2015,grijs2017}. Using such
middle-wavelength images guarantees that we do not reject extremely
blue or extremely red objects from our initial sample of cluster
candidates. The standard deviations ($\sigma_{\rm sky}$) of the number
of counts in empty sections of images in both filters were determined
using the {\sc Iraf\footnote{The Image Reduction and Analysis Facility
    ({\sc Iraf}) is distributed by the National Optical Astronomy
    Observatories, which is operated by the Association of
    Universities for Research in Astronomy, Inc., under cooperative
    agreement with the U.S. National Science Foundation. We used {\sc
      Iraf} version 2.16.1 (October 2013) for the data reduction
    performed in this study.}/imstat} routine. Multiples of this
background count in units of $\sigma_{\rm sky}$, i.e. [2, 3, 4, 5,
  6]$\sigma_{\rm sky}$, were used as thresholds above which the
numbers of source detections in both filters were calculated with the
help of the {\sc idl/find} task. The number of detections initially
decreases rapidly with increasing detection threshold; subsequently,
the curves become shallower. This suggests that our source detections
are dominated by noise in the low-threshold regime
\citep[e.g.][]{Barker2008}. When the rapid decline changes to a
flatter trend, `real' objects (either cluster candidates or
background-intensity variations) start to dominate our
detections. Therefore, we need to determine the knee in the curve,
where real sources begin to dominate the noise. We found that the most
suitable minimum detection thresholds are at 3$\sigma_{\rm sky}$ or
4$\sigma_{\rm sky}$ depending on the adopted master filters and the
specific ring. Note that by adopting these thresholds, we opted to set
{\it conservative} minimum flux limits; in this first selection step,
we did not want to discard objects that might be real clusters, but we
simply wanted to remove most of the spurious noise detections from our
sample; this procedure is directly based on that adopted and justified
by \citet[][]{deGrijs2013a,grijs2017}. Thus, in the remainder of this
paper, we will only consider the objects in the `source-dominated'
domain, i.e. for source brightnesses above the relevant detection
threshold.

To ensure that we are dealing with real objects in at least these two
middle-wavelength `master' filters, we applied a cross-identification
procedure in {\sc idl} to select only those sources that have
intensity peaks within 1.4 pixels of each other in our master filter
combination (i.e. allowing a maximum positional mismatch of only 1
pixel in both spatial dimensions). As shown by \citet[][]{grijs2017},
releasing this constraint, adjusted to 2- and 3-pixel positional
mismatches in both directions, does not have a significant effect on
the number of objects retained for further analysis. In our next step,
we made use of a standard Gauss-fitting routine in {\sc idl}, applied
to all selected sources, to determine their sizes, $\sigma_{\rm
  G}$. Size selection can help us to remove unlikely cluster
candidates. To define the minimum size for extended cluster
candidates, we generated artificial {\sl HST} point-spread functions
(PSFs) using the {\sc TinyTim} package \citep[][]{krist1997}, and
determined their best-fitting Gaussian widths. Cluster candidates with
sizes smaller than the model size (which may be artifacts of the
detector or caused by cosmic rays) were discarded. We adopted a
conservative size-cut criterion in order not to reject some marginally
extended sources. Although the series of selection criteria adopted
tends to lead to the rejection of a large fraction of our initial
source population, the method has been extensively validated and shown
to lead to well-understood final cluster samples
\citep[e.g.][]{anders2004b,Barker2008,deGrijs2013a,grijs2017}.

We next proceeded to obtain photometric measurements for our final
selection of sample clusters to acquire their SEDs. Our customized
{\sc idl} aperture-photometry task uses source radii and sky annuli
tailored to the properties of individual cluster candidates
\citep[e.g.][]{grijs2012,deGrijs2013a,li2015,grijs2017}. We used a
source aperture radius of $3 \sigma_{\rm G}$, and $3.5\sigma_{\rm G}$
and $5\sigma_{\rm G}$ for the inner and outer sky annuli,
respectively. The exact scaling was determined by checking the stellar
growth curves, to identify where the objects' radial profiles vanish
into the background noise. We confirmed that the choice of our source
radii was conservative enough so as not to miss any genuine source
flux, and we also verified that our selected sky annuli were not
significantly contaminated by neighboring sources. We calibrated our
aperture photometry using the prevailing zero-point offsets. The
photometric zero points (zpt) rely on the {\sl HST} image header
keywords {\sc photflam} and {\sc photzpt}, with zpt = $-$2.5 log({\sc
  photflam}) + {\sc photzpt}. We have included the multi-passband
cluster photometry for all sample galaxies except NGC 1512 and NGC
6951 in the Appendix; the equivalent data for NGC 1512 and NGC 6951
were published as Supplementary Data by de Grijs et al. (2017).

\subsection{Cluster parameter determination}

In order to determine the age ($\tau$), mass ($m_{\star}$), and
internal extinction ($A_{V}$) for each ring cluster candidate, we
compared the observed cluster SEDs to those of simple stellar
population (SSP) models using a $\chi^2$ minimization method
\citep[][]{anders2004a}:
\begin{equation}
\chi^2(\tau, A_{V}, m_{\star})=\sum_{N} \frac{({M_{\rm obs}-M_{\rm model}})^2}{\sigma_{\rm obs}^{2}}, 
\end{equation}
where $N$ is the number of available filters for each cluster, $M_{\rm
  obs}$ and $M_{\rm model}$ are the absolute magnitudes in each band,
respectively, and $\sigma_{\rm obs}$ are the observational
uncertainties. We used the {\sc galev} SSP models \citep[][and
  references therein; http://www.galev.org]{kotu09} to generate our
model SED suite, covering ages from $4 \times 10^6$ yr to $1.6 \times
10^{10}$ yr, with an age resolution of $4 \times 10^6$ yr. The model
grid was completed by inclusion of extinction effects
\citep[][]{cal2000}, with the reddening, $E(B-V)$, spanning the range
from 0.0 mag to 2.0 mag, with a resolution of 0.1 mag. We adopted a
\citet[][]{kroupa2001} stellar initial mass function (IMF) for stellar
masses from 0.1 $M_{\odot}$ to 100 $M_{\odot}$, which is the same as
used in Eq. (1) for determination of the SFR. Note that the broad-band
SED {\it shape} reveals information pertaining to a cluster's
best-fitting age, metallicity, and extinction, while the associated
cluster mass is the normalization factor to scale the models to the
observed SEDs. Each model SED (and its associated physical parameters)
was assigned a probability based on the $\chi^2$ value of this
comparison. For each cluster, we selected the model with the smallest
reduced $\chi^2$ to retrieve the cluster's most appropriate age, mass,
and extinction. Models with decreasing probabilities were summed up
until a 68.26\% total probability (1$\sigma$ confidence level) was
reached, to estimate the uncertainties in the best-fitting model
\citep[see][]{anders2004a}.

To determine robust ages and masses within a given cluster system, our
fitting algorithm requires cluster photometry in at least four
separate filters as input parameters \citep{grijs2005}, of which the
$U$ band filter is fundamental to break the age--extinction degeneracy
in the colors of young clusters \citep[][]{anders2004a}. In principle,
we need the wavelength coverage to be as broad as possible, but the
availability of {\sl HST} images limits the filter set used for our
SED fitting. \citet[][]{allard2006} and \citet[][]{sarzi2007}
demonstrated that the circumnuclear regions of barred massive
galaxies can be modeled well by adopting solar metallicity. Given
  that all galaxies in our sample are nearby luminous and massive
  spiral galaxies, we decided to keep the model cluster metallicities
unchanged by adopting solar metallicity, retaining the cluster ages,
masses, and extinction values as free parameters. The advantage of
this approach is that it leaves us with one fewer free parameter to
determine, which in turn renders our resulting age, mass, and
extinction estimates more reliable (we only need a minimum of three
filters to obtain credible cluster parameters).

As an example, Fig. 3 shows the distribution of our detected NGC 1512
cluster candidates in the age--mass plane. We have not included the
error bars in age and mass for the sake of clarity; the 1$\sigma$ age
and mass ranges are included in Table 2. The red points represent
clusters located in the nuclear ring region, while the blue points
represent the clusters found in the main nuclear disk. For this
galaxy, we have access to {\sl HST} images taken through the WFC3/UVIS
F336W, F438W, F555W, and F814W broad-band filters. Since our object
selection is based on source detection and cross-identification in the
F438W and F555W filters, our sample is intrinsically limited by the
least sensitive of the latter filters. Our NGC 1512 cluster sample is
therefore a `$B$-band detection-limited' sample. As regards the
procedures used to perform our completeness tests, we refer the reader
to \citet[][their Section 2.5]{grijs2017}, where a detailed
description can be found. Here, for the sake of clarity, we
  briefly describe the main steps followed for determining the
  completeness limits. For a given filter, we generated 500 artificial
  clusters (with sizes representative of our target galaxy's cluster
  population; see below) by applying standard {\sc Iraf} tasks, which
  we subsequently added to the ring region of our science image,
  adopting randomly generated ($x, y$) coordinates. We assigned the
  mean size of the galaxy's genuine cluster sample to our artificial
  clusters. The input artificial sources, as well as their photometry
  in the relevant passbands, were retrieved using the same source
  discovery routines as applied to find the real clusters, in order to
  evaluate how many input artificial objects may have been missed by
  our processing technique. Finally, we repeated this procedure by
  varying magnitudes of the input sources from 19.0 mag to 27.0 mag,
  in steps of 0.5 mag, in all relevant filters.
 
The black solid line in Fig. 3 is the evolutionary track of an SSP
model for the $B$ band's 90\% completeness limit, $m_{\rm
  F438W}^{90\%} = 22.7$ mag, at the distance of NGC 1512. This mass
evolution curve was calculated using the {\sc galev} models for solar
metallicity and zero extinction, i.e., the completeness limit of 22.7
mag in the F438W passband was converted to a minimum mass estimate for
each age. This shows the expected effect of evolutionary fading of an
instantaneously formed SSP. As one can see from Fig. 3, there are some
clusters with best-fitting masses below the limiting model
prediction. This is probably due to the fact that the mass of a
cluster is determined by scaling the complete model SED, rather than
scaling a single band, which in essence implies that stochastic
sampling effects might cause the observed scatter in cluster mass.


\begin{table}
\begin{center}
\caption{Derived NGC 1512 ring star cluster properties}
\begin{tabular}{cccccccccc}
\hline
\#&\multicolumn{3}{c}{log($t$ $\rm{yr}^{-1}$)}& &\multicolumn{3}{c}{log($M_{\rm cl} / M_{\odot}$)} & &$E(B-V)$\\
\cline{2-4} \cline{6-8}
\colhead{}&\colhead{Best}&\colhead{Min.}&\colhead{Max.}&\colhead{}&\colhead{Best}&\colhead{Min.}&\colhead{Max.}&\colhead{}&\colhead{(mag)}\\
\hline
0&9.00&6.60&9.58&&5.67&4.80&6.28& &0.0\\
1&8.23&6.60&8.99&&4.88&3.47&5.38& &0.1\\
2&6.60&6.60&9.58&&5.06&4.37&6.35& &0.8\\
3&9.06&6.60&10.18&&5.18&3.35&6.22& &0.0\\
4&6.60&6.60&9.20&&4.03&3.16&5.30& &0.6  \\
$\cdots$ & $\cdots$ & $\cdots$ & $\cdots$ && $\cdots$ & $\cdots$ & $\cdots$ & & $\cdots$ \\
\hline
\end{tabular}
\end{center}
\tablecomments{Table 2 is published in its entirety in the electronic
  version of {\it The Astrophysical Journal}. A portion is shown here
  for guidance regarding its form and content.}
\end{table}

Most of the ring clusters are matched by model SEDs that are younger
than 100 Myr, covering a continuous age range, and they have low
extinction values, on average $E(B-V)$ = 0.18 mag, which is in good
agreement with \citet[][]{maoz2001}. The figure also exhibits a number
of small-scale features in the cluster distribution, with gaps at some
ages and apparent overdensities of clusters at other ages. These types
of features are expected when estimating ages by comparing observed
cluster SEDs with SSP models provided for a distinct set of ages. We
inspected an apparent overdensity of data points at $\log(t \mbox{
  yr}^{-1}) = 6.6$ and found that 90 ring clusters are located in this
`chimney' feature. This artifact is attributed to our fitting
routines, i.e. the large density of clusters at this location is
simply caused by the fact that our youngest isochrone has an age of 4
Myr (we are limited by the age range covered by the Padova isochrones
on which the {\sc galev} SSP models are based). Younger clusters would
therefore be assigned the minimum model age
\citep[cf.][]{bastian2005,gieles2005}. Because of the limitations
related to the number of available filters, we are forced to use an
almost continuously distributed model suite to fit our observed SEDs
derived from observations in a number of discrete filters. This will
inevitably lead to local minima in the $\chi^2$ landscape (manifested
as older chimneys in Fig. 3). In fact, we would expect their ages to
scatter somewhat more (by about 0.15 dex) around the mean age of such
a chimney \citep[e.g.][]{anders2004a}. However, these small-scale
artifacts do not significantly affect the broad distribution of
cluster masses and ages. The ring cluster parameters for the other
galaxies in our sample are included in the Appendix. In Table 3,
  we present the basic parameters of our sample ring cluster
  populations. 
\begin{figure}[ht!]
\begin{center}
\includegraphics[width=1.1\columnwidth]{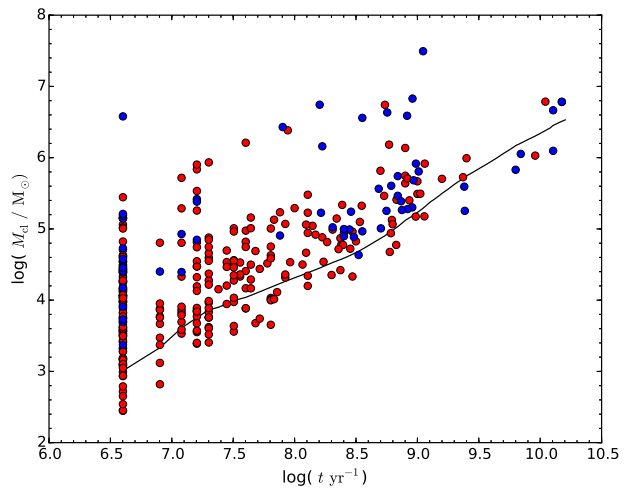}
\caption{NGC 1512 star cluster population in the age--mass
  plane. Clusters identified by red points are associated with the
  circumnuclear starburst ring, while the blue points indicate
  clusters found in the main galactic disk. The solid black line
  represents the cluster mass limit as a function of age estimated
  from the evolutionary tracks assuming a 90\% completeness limit of
  22.7 mag in the $B$ band, $m_{\rm F438W}^{90\%} = 22.7$ mag, for
  solar metallicity and zero extinction. }
\label{}
\end{center}
\end{figure}

\begin{table*}
\begin{threeparttable}
\begin{center}
\begin{minipage}{160mm}
\caption{ Basic parameters of the ring star cluster populations}
\label{comparison.tab}
\begin{tabular}{lcccccccc}
\hline
\hline \\[-1.4ex]
\colhead{Galaxy}&\colhead{$N$}&\colhead{$f$}&\colhead{$ \langle \Sigma
  \rangle$ (kpc$^{-2}$)}&Limiting filter&Mag&$\langle$log($t\,\,{\rm
  yr}^{-1}$)$\rangle$&$\langle$log($M_{\rm
  cl}$/$M_{\odot}$)$\rangle$&log$(M_{\rm YMC}^{\rm
  tot}/M_{\odot})$\\[0.8ex]
\,\,\,\,\,\,\,\,\,\,\,(1)&(2)&(3)&(4)&(5)&(6)&(7)&(8)&(9)\\[0.6ex]
\hline\\[-1.4ex]

ESO 565-11&143&0.77&8&F439W&22.4&7.59&5.56&6.7 \\
 IC 342&64&0.17&1280&F438W&20.5&8.20&4.73&6.0\\
NGC 1097&1064&0.55&213&F438W&22.3&8.83&6.11&7.4 \\
NGC 1326&242&0.56&484&F439W&21.3&8.11&5.58&6.6 \\
NGC 1512&319&0.33&363&F438W&22.7&8.18&5.15&5.9 \\
NGC 1672&302&0.30&3020&F435W&21.3&8.69&6.25&6.8\\
NGC 2997&251&0.57&179&F555W&24.1&7.72&4.41&5.4 \\
NGC 3081&27&0.63&18&F439W&21.3 &7.87&6.39&6.5 \\
NGC 3351&448&0.59&1120&F438W&21.0 &8.47&5.40&6.4 \\
NGC 4303&143&0.78&600&F555W&22.3&7.79&5.10&6.2\\
NGC 4314&209&0.82&565&F439W&23.0 &7.96&4.60&5.8\\
NGC 6782&45&0.92&7&F450W&20.4&8.59&5.92&6.4\\
NGC 6951&82&0.79&108&F547M&21.8&8.24&6.08&6.6 \\
NGC 7217&116&0.64&49&F450W&23.6&8.64&5.32&5.7\\
NGC 7469&57&0.81&61&F435W&21.0 &8.63&7.11&7.4\\
NGC 7742&271&0.61&22&F555W&23.7 &7.44&5.49&7.2\\
UGC 3789&81&0.77&39&F438W&22.6  &7.52&5.93&6.7\\
\hline

\end{tabular}
\begin{tablenotes}
\item { Notes. Columns: 1 -- Host galaxy name; 2 -- Number of
  clusters in the final selected ring cluster population; 3 -- Ratio
  of the number of objects in the final ring cluster population with
  respect to that in the initial cluster population; 4 -- Ring cluster
  surface density in units of kpc$^{-2}$; 5, 6 -- Limiting master filter used for the cluster selection and corresponding
  90\% completeness limit; 7 -- Average age of the ring cluster
  sample; 8 -- Average mass of the ring cluster sample; 9 -- Total
  mass of YMCs in the ring region with ages up to 10 Myr and masses greater than $10^4 M_{\odot}$. }
\end{tablenotes}
\end{minipage}
\end{center}
\end{threeparttable}
\end{table*}

\subsection{Determination of the integrated ring properties}

To measure the physical properties of nuclear rings, we follow exactly
the same procedures as previously established in Paper I, where a
detailed description can be found. We refer the reader to that paper
for details, and so we only summarize the main steps here. Based on
the observed radial profile shape of the nuclear region (see Fig. 2 of
Paper I), we divide the images into two parts, i.e. the host galaxy's
background and nuclear ring areas. We model these galactic components
by adopting analytical functions, using the {\sc galfit} fitting
program, including a truncation function for the nuclear ring, and a
combination of Gaussian and S\'{e}rsic functions aiming for the
background galactic disk \citep[for a detailed description, see
  also][]{peng2010}. We found a good match between the best-fitting
model and the real data, and we demonstrated that the results are not
affected by the lower {\sl Spitzer} resolution. We applied our
improved method to our ring sample to derive their integrated
optical-to-mid-infrared SEDs.

We next fitted the observed SEDs of our ring sample to determine their
average age and total stellar mass using Flexible Stellar Population
Synthesis models \citep[FSPS;][]{conroy2009,conroy2010}. By assuming a
Kroupa (2003) IMF, stellar population model templates were produced,
with the metallicity again fixed to the solar value. The age--mass
distribution of our ring sample is shown in Fig. 4, which reveals an
average age of $\sim$1.2 Gyr. Both previous observational
\citep[e.g.][]{allard2006,sarzi2007} and numerical
\citep[e.g.][]{regan2003} results indicate that star-forming nuclear
rings are long-lived structures with multiple epochs of starburst
activity. Based on the observed nuclear ring fraction ($\sim20\pm2$\%)
in disk galaxies from their unbiased sample, \citet[][]{comeron2010}
estimated an approximate effective nuclear ring lifetime on the order
of 2--3 Gyr, which is indeed compatible with our sample-averaged
current ring age. Nevertheless, we note that our measurements are much
more accurate given our systematic approach based on multi-passband
SED analysis. The derived ring masses for our sample galaxies vary
widely; they are in the range of $\sim 10^7-10^9\, M_{\odot}$, while
the total gas mass in each ring is almost constant at approximately a
few $\times 10^8 \, M_{\odot}$
\citep[e.g.][]{buta2000,benedict2002,sheth2005}. The ages and masses
we determined for our nuclear rings, as well as their 1$\sigma$
uncertainties, are summarized in columns 2 and 3 of Table 4,
respectively. For rings with measured SFRs (see Table 4, column 4),
the best-fitting ring stellar masses are an order of magnitude higher
than the deduced stellar masses (based on assuming constant SFRs as
observed over the current ring lifetimes, i.e., their ages). This
therefore illustrates that nuclear ring star-formation histories are
more complicated than represented by our simple assumption of a
constant SFR. Indeed, some evidence has revealed that the observed
emission lines in starburst rings might be best modeled by adopting
multiple starburst episodes of varying intensity rather than by a
constant SFR \citep[e.g.][]{allard2006,sarzi2007}

\begin{table*}
\begin{threeparttable}
\begin{center}
\begin{minipage}{130mm}
\caption{Integrated nuclear ring properties derived in this paper}
\label{comparison.tab}
\begin{tabular}{lcccc}
\hline
\hline \\[-1.4ex]
\colhead{Galaxy}&\colhead{log($t\,{\rm yr}^{-1}$)}&\colhead{log($M$/$M_{\odot}$) }&\colhead{SFR ($M_{\odot}$ yr$^{-1}$)}&$\Sigma_{\rm SFR}\,(M_{\odot}\,{\rm yr}^{-1}\,{\rm kpc}^{-1})$\\[0.8ex]
\hline\\[-1.4ex]
ESO 565-11&$8.99_{-2.3}^{+0.06}$&$8.96_{-0.31}^{+0.15}$&---&---\\[1ex]
IC 342&$9.71_{-0.96}^{+0.28}$&$7.99_{-0.17}^{+0.16}$&---&---\\[1ex]
NGC 1097&$8.31_{-0.25}^{+1.22}$&$9.93_{-0.14}^{+0.05}$&$2.307\pm0.4$&$0.468\pm0.09$\\[1ex]
NGC 1326&$9.32_{-0.97}^{+0.25}$&$8.97_{-0.1}^{+0.06}$&---&---\\[1ex]
NGC 1512&$7.63_{-0.31}^{+0.15}$&$7.13_{-0.11}^{+0.11}$&$0.08\pm0.01$&$0.09\pm0.012$\\[1ex]
NGC 1672&$9.65_{-0.54}^{+0.23}$&$9.67_{-0.17}^{+0.1}$&---&---\\[1ex]
NGC 2997&$8.32_{-0.23}^{+1.36}$&$8.85_{-0.26}^{+0.2}$&---&---\\[1ex]
NGC 3081&$9.01_{-0.66}^{+0.47}$&$8.77_{-0.22}^{+0.07}$&---&---\\[1ex]
NGC 3351&$8.38_{-0.06}^{+0.07}$&$9.1_{-0.06}^{+0.07}$&$0.294\pm0.033$&$0.709\pm.081$\\[1ex]
NGC 4303&$9.51_{-0.6}^{+0.33}$&$8.75_{-0.17}^{+0.17}$&---&---\\[1ex]
NGC 4314&$7.58_{-0.18}^{+0.17}$&$7.51_{-0.11}^{+0.12}$&$0.04\pm0.006$&$0.109\pm0.017$\\[1ex]
NGC 6782&$7.96_{-1.15}^{+1.32}$&$9.44_{-0.19}^{+0.16}$&---&---\\[1ex]
NGC 6951&$8.61_{-0.11}^{+1.08}$&$9.4_{-0.26}^{+0.11}$&$0.188\pm0.035$&$0.248\pm0.046$\\[1ex]
NGC 7217&$7.93_{-0.18}^{+1.51}$&$8.39_{-0.21}^{+0.15}$&$0.065\pm0.013$&$0.027\pm0.005$\\[1ex]
NGC 7469&$9.16_{-3.58}^{+0.65}$&$9.66_{-0.31}^{+0.20}$&---&---\\[1ex]
NGC 7742&$8.59_{-1.98}^{+0.10}$&$9.77_{-0.13}^{+0.09}$&---&---\\[1ex]
UGC 3789&$7.88_{-0.31}^{+0.33}$&$9.06_{-0.1}^{+0.1}$&---&---\\[1ex]
\hline
\end{tabular}
\begin{tablenotes}
\item {\bf Notes}. Columns: 1 -- host galaxy name; 2 -- best-fitting
  age; 3 -- total stellar mass; 4 -- current SFR for the past 10 Myr
  in units of $M_{\odot}$ yr$^{-1}$; 5 -- corresponding SFR surface
  density in units of $M_{\odot}$ yr$^{-1}$ kpc$^{-1}$.
\end{tablenotes}
\end{minipage}
\end{center}
\end{threeparttable}
\end{table*}

\section{Analysis and discussion of possible correlations}

\begin{figure}[ht!]
\begin{center}
\includegraphics[width=1.1\columnwidth]{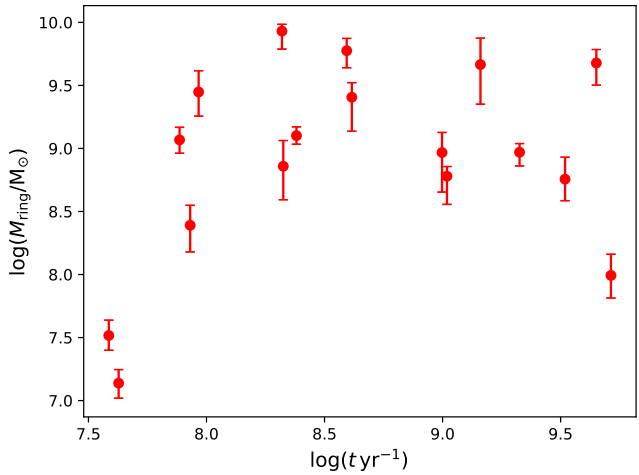}
\caption{Average ring age versus total stellar mass in the ring, both
  on logarithmic scales. The associated error bars are the 16$^{\rm
    th}$ and 84$^{\rm th}$ percentiles of the model posteriors. The
  average ring age is around 1.2 Gyr.}
\label{}
\end{center}
\end{figure}

\begin{figure}[ht!]
\begin{center}
\includegraphics[width=1.1\columnwidth]{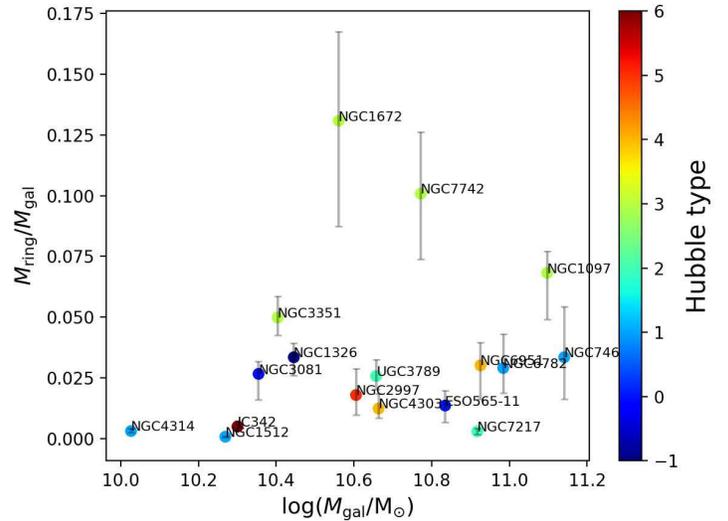}
\caption{Ring mass fraction with respect to the total stellar mass of
  the host galaxy, as a function of galaxy mass (logarithmic
  scale). The color coding reflects the galaxies' the revised Hubble
  types, which range from S$0^+$($T=-1$) to Scd ($T=6$) for our galaxy
  sample.}
\label{}
\end{center}
\end{figure}

In this section we will explore the possible correlations between a
range of integrated ring and host galaxy properties, as well as those
pertaining to the ring cluster populations. Figure 5 displays $M_{\rm
  ring}/M_{\rm gal}$, i.e. the ratio of the ring mass to the overall
host galaxy's stellar mass, as a function of $\log(M_{\rm gal})$,
color-coded based on the galaxies' revised Hubble types. The stellar
masses were calculated based on the color-dependent $K$-band
mass-to-light ratios ($M_{\rm gal}/L$), using \citep{bell2003}
\begin{equation}
{\log}(M_{\rm gal}/L)= a_{k}+b_{k}\times(B-V),
\end{equation}
where the $M_{\rm gal}/L$ ratio is given in solar units. The $K$-band
luminosity ($L$) and the corresponding $(B-V)$ color were obtained
from the HyperLeda\footnote{http://leda.univ-lyon1.fr} data base. Of
particular interest in this figure is that all Sb-type ($T=3$)
galaxies, i.e. NGC 1097, NGC 1672, NGC 3351, and NGC 7742, have the
highest stellar ring-mass fraction with respect to the stellar mass of
their host galaxies.

\subsection{Star formation in the ring}

For six of the 17 galaxies in our sample (including NGC 1512, NGC
1097, NGC 3351, NGC 4314, NGC 6951, and NGC 7217), we could calculate
the current ring SFR, since both H$\alpha$ and 8 $\mu$m images were
available (see Table 4). Figure 6 shows the ring star-formation
properties, SFR and $\Sigma_{\rm SFR}$, compared with various ring
parameters; $\Sigma_{\rm SFR}$ is the SFR surface density, defined as
the SFR divided by the ring area. In the top row of Fig. 6, we show
how $\Sigma_{\rm SFR}$ and SFR are correlated with the ring mass
fraction (left) and the absolute stellar mass in the ring (right).
 The top right-hand panel is reminiscent of the main sequence (MS)
  of star-forming galaxies \citep[e.g.][]{renzini2015,lin2017}, but it
  applies here only to local starburst-ring environments within
  galaxies. Interestingly, these subgalactic regions exhibit a similar
  correlation as their full counterparts, except for NGC 1512 and NGC
  4314, which are also the two least massive rings with the youngest
  ages in our sample. The blue line in this panel denotes the
  $\log(M_{\star})$--log(SFR) correlation for MS star-forming
  galaxies, characterized by a best-fitting slope of 0.8
  \citep[e.g.,][]{speagle2014,pannella2015}. The left-hand panel in
  the second row of Fig. 6 shows a negative trend for the specific SFR
  (i.e. the ring stellar mass-normalized SFR, SFR/$M_{\rm ring}$) as
  function of the ring mass. Again, this trend is also seen for the
integrated properties of local galaxies
\citep[e.g.][]{behroozi2013}. The right-hand panel in the second row
explores the YMC mass fraction as a function of $\Sigma_{\rm SFR}$,
 with the YMCs' total masses listed in column (9) of Table 3. To
obtain the ring YMC samples, we selected clusters younger than 10 Myr
and more massive than $10^4$ $M_{\odot}$, but this exploration does
not reveal any evident trend.

The two panels in the third row of Fig. 6 show the dependence of the
non-axisymmetric torque parameter $Q_{\rm g}$ to the SFR and
$\Sigma_{\rm SFR}$, respectively, with both panels demonstrating
striking correlations between star-forming properties of rings and the
non-axisymmetric perturbations by host galaxies. As the bar strength
increases, the star-forming intensity in the ring is significantly
weakened, except for one discrepant data point (NGC 7217). Unlike the
other sample galaxies, NGC 7217 is an unbarred galaxy containing a
counter-rotating component. Note that strongly barred galaxies tend to
have low SFRs in their rings \citep{mazzuca2008,comeron2010}. It thus
appears that our observed trends are in accordance with previously
published results, although we emphasize that our sample may be
suffering from unquantified selection effects. Examining the
  bottom left-hand panel of Fig. 6, we also note that the rings'
  specific SFR shows a weakly increasing trend with increasing $Q_{\rm
    g}$. These observational results may be in contrast with the
general expectation that strong bars will tend to drive more material
to the central regions of their host galaxies and thus lead to rings
with higher SFRs. However, we note that nuclear rings are the products
of a complicated interplay of multiple factors, and it is likely that
other physical determinants, such as the SFR history, the gas density
in the ring region and the gas content within the bar radius, may also
have an impact. We leave an exploration of such effects for a
follow-up study.

\begin{figure}[ht!]
\begin{center}
\includegraphics[width=1.1\columnwidth,height=105mm]{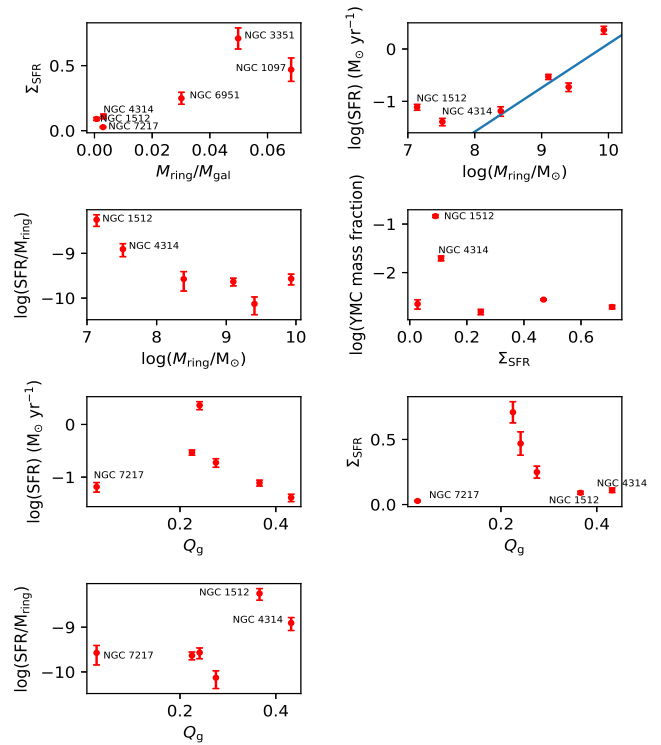}
\caption{Ring star-forming properties versus a number of ring
  parameters. (Top left) Ring SFR surface density, $\Sigma_{\rm SFR}$,
  as a function of $M_{\rm ring}$/$M_{\rm gal}$; (top right) SFR in
  logarithmic units of $M_{\odot}$ yr$^{-1}$ as a function of ring
  mass, equivalent to a `localized' galaxy main sequence. The
    blue line shows the canonical slope of 0.8 for MS star-forming
    galaxies; (second row, left) Specific SFR (sSFR) normalized by and
    as a function of ring mass; (second row, right) YMC mass fraction
    with respect to ring mass versus $\Sigma_{\rm SFR}$. (third row)
    Dependence of $Q_{\rm g}$ on the SFR (left) and $\Sigma_{\rm SFR}$
    (right). The panel in the bottom-left corner represents a weakly
    increasing trend of ring sSFR and $Q_{\rm g}$.}
\label{ }
\end{center}
\end{figure}

\subsection{Nuclear rings and young star clusters}

Since bars play an important role in nuclear ring formation and their
evolution, the bar strength is possibly the dominant factor in
determining the properties of the newly formed ring cluster
population. In Fig. 7, we plot the ring's YMC mass fractions with
respect to the ring masses as a function of $Q_{\rm g}$. For the ring
YMCs, we used the same selection criteria as above. Two galaxies,
  NGC 7217 and NGC 7742, which are unbarred and show minor-merger
  evidence are labeled in yellow. There is no apparent relationship
between the integrated ring properties and the YMC populations.

\begin{figure}[ht!]
\begin{center}
\includegraphics[width=1.1\columnwidth]{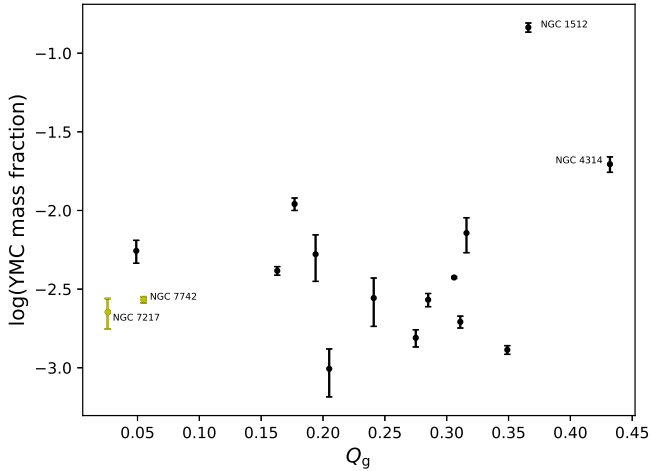}
\caption{YMC mass fractions with respect to the ring masses as a
  function of $Q_{\rm g}$. The selected YMCs, located in the ring
  regions, are all younger than 10 Myr and more massive than $10^4
  M_{\odot}$.  The two unbarred galaxies NGC 7217 and NGC 7742
    which exhibit minor-merger evidence are marked in yellow.}
\label{}
\end{center}
\end{figure}

Figure 8 shows the YMC mass fractions as a function of the integrated
ring stellar masses (left) and ages (right), where the rings of
  the unbarred galaxies in our sample (i.e., NGC 7217, NGC 7742, and
  UGC 3789) are again marked in yellow. In the left-hand panel, a
pronounced trend is found, irrespective of the presence of the
  unbarred sample galaxies, showing that the YMC mass fraction
decreases rapidly as the host galaxy's ring becomes more massive. Our
ring masses derived from broad-band SED fitting span a large range of
more than two orders of magnitude, from $1.37\times 10^7\,M_{\odot}$
for NGC 1512 to $8.5\times 10^9\,M_{\odot}$ for NGC 1097. YMC samples
with ages of up to $\sim$ 10 Myr nevertheless account for merely small
fractions of the total ring masses. This is why more massive rings
tend to have lower YMC fractions. In the right-hand panel of Fig. 8,
where we compare the YMC fractions with the average ring ages, there
is no noticeable relationship with or without rings of unbarred
  galaxies. Considering the broad age and mass ranges among our
sample of rings, as well as the wide range of non-axisymmetric
parameters ($Q_{\rm g}$) of their host galaxies, our ring catalog is
reasonably representative of nearby nuclear-ring galaxies in general,
despite the relatively small number of rings. Interestingly, the
nuclear ring of NGC 1512 is the second youngest ($\sim$40 Myr-old)
ring in the sample, with the lowest stellar mass, but it is
characterized by the highest YMC mass fraction when normalized by
parent ring stellar mass, as shown in Fig. 8.

\begin{figure}[ht!]
\begin{center}
\includegraphics[width=1.1\columnwidth]{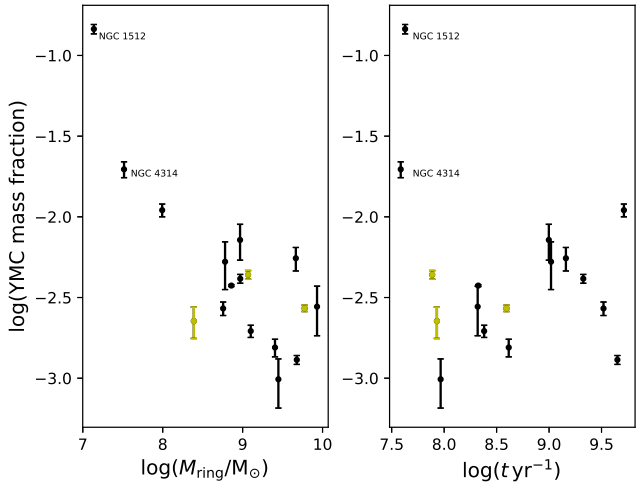}
\caption{YMC mass fractions as a function of total ring stellar masses
  (left) and best-fitting ages (right), respectively. The three
    yellow points represent the rings in our unbarred sample galaxies,
    NGC 7217, NGC 7742, and UGC 3789.}
\label{}
\end{center}
\end{figure}

 Figures 6 to 8, show that the ring properties in NGC 1512 and NGC
  4314 are outliers in almost all cases. These are the least massive
  and youngest rings with the highest specific SFRs and YMC mass
  fractions (for which the relatively lower ring stellar masses may
  play a crucial role in determining their physical properties). We
  have marked these two rings separately in Figs 6--8. These two
  youngest rings are different from the other rings in our sample, and
  excluding them will affect the above observed trends and correlation
  to a large extent. For example, the decreasing relation in the
  left-hand panel of Fig. 8 would be less significant if we do not
  consider the two youngest rings, although we concede that the
  statistics basis of our sample is limited. The same situation also
  applies to both Fig. 7 and the relevant panels of Fig. 6. While the
  most likely interpretation of the physical reasons underlying any of
  the trends may be ambiguous, perhaps the very young ring populations
  actually bias the overall picture. Alternatively, the complex SFHs
  operating in the nuclear rings may affect the results differently.

As regards Fig. 4, even though the number of objects in our sample is
small, it appears that the age--mass distribution of our ring sample
represents an evolutionary signature, at least qualitatively, with the
least massive rings (NGC 1512 and NGC 4314) occupying the youngest end
of the rings' evolutionary sequence. Meanwhile, these two special
rings deviate significantly from the star-forming galaxy MS, as shown
in the top right-hand panel of Fig. 6. We therefore propose that they
may represent the young-age tail of the nuclear ring age distribution,
rather than forming a separate population of nuclear rings. As the
ring ages increase, more gaseous material is channeled into the ring
region by the galactic bar potential, which contributes to the gradual
increase in the total stellar mass of the nuclear rings. However, we
caution that this is only a tentative explanation. The physical
mechanisms behind the observed phenomena are difficult to quantify
without knowing the details of the ring SFHs, gas content, and
dynamics in the circumnuclear regions, among others, all of which
should be further explored in future studies. A larger ring sample
benefiting from observations of higher spatial resolution and a longer
wavelength coverage would also help to further resolve these
conundrums. This is left for future work.

A number of early studies \citep[e.g.][]{grijs2012,kon2013,vaisa2014}
have revealed an environmental dependence of the timescale of young
cluster disruption based on analyses of cluster mass functions (CMFs).
CMFs are basic tools to explore the formation and evolution histories
of entire star cluster populations, as well as of the imprints the
galactic environment leaves on the properties of its cluster
population. It is well-known that the distributions of young star
cluster populations can be approximated by a power-law function of the
form ${\rm d} N(M) \propto M^{-\alpha} {\rm d}M$ with $\alpha \sim 2$
for a large range of cluster masses in nearby spiral and starburst
galaxies \citep[e.g.,][]{zhang1999,grijs2003,port2010,fall2012}. It is
of paramount importance to understand what governs the cluster
formation process and how the ring environment may affect their
formation and evolution. A more systematic assessment of CMF evolution
in such rings therefore has the potential to open up this entire
field.

We next proceeded to explore the impact of various ring properties on
the form of the CMF defined by young ring clusters. In Fig. 9, we show
the ring CMFs of young star cluster populations, with ages younger
than 10 Myr. The associated error bars are derived from Poissonian
statistics in the respective mass bins. To the left of CMF turnover
(commonly defined as the conservative detection limits) in each panel,
these relatively low-mass star clusters are severely affected by
incompleteness effects. The vertical dashed lines in each panel are
the mass-detection limits calculated based on the corresponding 90\%
photometric completeness limit for the ring area only and for the most
restrictive filter (see also Section 3.2 for a detailed discussion),
and based on {\sc galev} SSPs.

Since the mass-to-light ratio increases with age, we adopted the
oldest age (10 Myr) to estimate the mass threshold. We can discern
clearly, for all galaxies, that the cluster mass thresholds for the
oldest age of 10 Myr are consistent with (or a little below) the
observed peaks in the CMFs. We therefore used a pure power-law
function to fit the data points at the high-mass ends, i.e. on the
right-hand side of the CMFs, which are observationally most complete.
The best-fitting indices $\alpha$ and the associated errors are also
included in each panel. The best-fitting power-laws are shown as blue
lines. We did not fit power-law functions to IC 342 or NGC 3081
because of the small numbers of clusters in those galaxies. Figure 10
shows the best-fitting power-law index as a function of $Q_{\rm
  g}$. It is not surprising that no tight relationship is observed
given the complicated interplay between the local star-forming
environments and the young star cluster populations. The bar strength
$Q_{\rm g}$ may just be one of the dominant factors (in addition to
the star-formation history, the gas content, the inflow rate,
dynamical effects, etc.) in determining the properties of the young
cluster systems in the rings. Further statistical analysis of larger
ring samples is required before we can make definitive statements.

\begin{figure*}[ht!]
\begin{center}
\includegraphics[width=1.8\columnwidth]{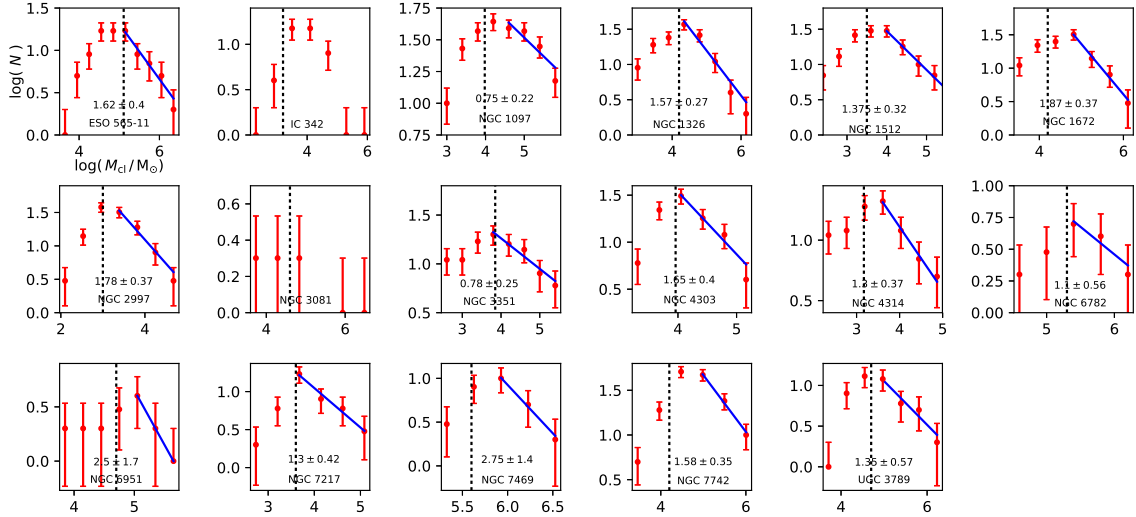}
\caption{CMFs of young star cluster populations, with ages younger
  than 10 Myr, are shown as red points. The error bars are Poissonian
  errors. The vertical black dashed lines highlight the 90\% detection
  limit in mass for the ring area based on the relevant limiting
  passband used for source detection and cross-identification. We
  adopted the oldest age to calculate the cluster mass thresholds,
  i.e. 10 Myr. As displayed by the blue lines, the high-mass ends of
  the CMFs which are much less affected by observational
  incompleteness are fitted by power laws, with the derived power-law
  slope $\alpha$ as well as fitting error noted in each panel (except
  for IC 342 and NGC 3081).}
\label{}
\end{center}
\end{figure*}

\begin{figure}[ht!]
\begin{center}
\includegraphics[width=1.1\columnwidth]{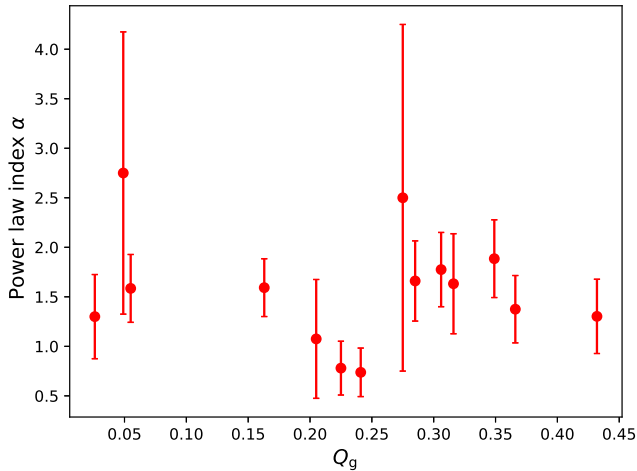}
\caption{Best-fitting power-law index $\alpha$ for the high-mass end
  of the young star cluster CMF as a function of $Q_{\rm g}$.  }
\label{}
\end{center}
\end{figure}

\section{Summary and Conclusions}

In this paper, we aimed to look for the possible correlations between
ring properties and their host galaxy parameters, and investigate the
influence of ring properties on their young star cluster systems.
Based on an extensive survey of the literature and imaging archives
(i.e. the Hubble Legacy Archive and the Spitzer Heritage Archive), we
collected a catalog of 17 nearby nuclear rings which were observed and
well-resolved in multiple passband by both the {\sl HST} and {\sl
  Spitzer}. Our sample is currently the most complete collection of
nuclear rings with high-resolution observations, which enables us to
simultaneously study the ring properties and their star cluster
populations. We applied our recently improved method to our ring
sample to derive the integrated ring properties, including their SEDs,
the average ages, and the total stellar masses, while for each ring we
also compiled catalogs of ring cluster populations with their ages and
masses estimated based on fitting their SEDs with SSP models. The
average ring age for the sample is around 1.2 Gyr, i.e. well within
the average ring lifetime proposed by \citet{comeron2010} based on the
observed nuclear ring fraction in disk galaxies. However, we emphasize
that our SED fit results are much more accurate than their rough
estimations. The rings' total stellar masses span a large range of
$\sim 10^7-10^9\,M_{\odot}$.

We found that Sb-type ($T = 3$) galaxies tend to have the highest
$M_{\rm ring}/M_{\rm gal}$ ratios. Six nuclear rings (NGC 1512, NGC
1097, NGC 3351, NGC 4314, NGC 6951, and NGC 7217) have observational
combinations of H$\alpha$ and 8 $\mu$m filters, allowing us to obtain
their current ring SFRs, i.e. within the last 10 Myr, based on the
prescription of \citet[][]{kennicutt2009}. It is clear that
$\Sigma_{\rm SFR}$ increases with $M_{\rm ring}/M_{\rm gal}$, and that
the ring SFRs are correlated with the rings' stellar masses, which is
reminiscent of the main sequence of star-forming galaxies, but only
for local starburst-ring regions in this case, with the two
  outlier galaxies NGC 1512 and NGC 4314 located slightly off-trend.
A clearly decreasing relationship was found between SFR/${\rm M}_{\rm
  ring}$ on the one hand and $M_{\rm ring}$ on the other, because the
ring stellar mass coverage in our sample is much more extensive than
their SFRs. We did not uncover any dependence of the YMC mass fraction
(with respect to the ring stellar mass) on $\Sigma_{\rm SFR}$,
probably due to small-number statistics. Excluding the outlier ring in
NGC 7217, there are significant correlations between star-forming
parameters (SFR and $\Sigma_{\rm SFR}$) and the non-axisymmetric
parameter $Q_{\rm g}$, which thus corroborates previous discoveries
that strongly barred galaxies tend to have lower SFRs in their nuclear
rings. However, we caution that our sample is limited. Other physical
effects may also play important roles in the complexities of the
star-formation properties, and compilation of a more extended ring
sample will help further address these fundamental questions. The
number of rings with measured SFRs in our sample is too limited to
make conclusive statements. We also found that the ring stellar masses
resulting from SED fitting are much higher than the values derived
based on assuming a constant current SFR over a ring's lifetime,
indicating that the star-forming history of nuclear ring is likely
more complex.

 Two special rings in NGC 1512 and NGC 4314 are usually located
  away from the trends observed for the other rings in our sample. We
  speculate that they may represent the young extremity of the nuclear
  ring age distribution rather than a separate class of nuclear
  rings. We explored the correlations, if any, between bar strength
and YMC mass fraction for clusters more massive than $10^4 M_{\odot}$
and with ages younger than 10 Myr, although no clear trends were
found. Since our sample covers a large mass range, as inferred from
SED analysis, rings with higher stellar masses tend to be associated
with lower YMC mass fractions, and high-mass rings appear to exhibit
significant scatter in the distribution of the YMC mass
fractions. These trends change little even if we exclude the two
  youngest rings in our sample (NGC 1512 and NGC 4314). This does not
apply to the ring ages. Finally, we analyzed the CMFs of the YMC
populations for all rings and fitted the high-mass regime, which
suffers less from incompleteness effects, using the canonical
power-law function. The best-fitting power-law indices were compared
with the host galaxies' bar strengths, but no correlations were found.

\section*{Acknowledgments}

We are greatly indebted to Peter Anders for his help with the SED
fitting techniques. This paper is based on observations obtained with
the NASA/ESA {\sl HST} and obtained from the Hubble Legacy Archive,
which is a collaboration between the Space Telescope Science Institute
(STScI/NASA), the Space Telescope European Coordinating Facility
(ST-ECF/ESA), and the Canadian Astronomy Data Centre
(CADC/NRC/CSA). This research has also made use of NASA's Astrophysics
Data System Abstract Service. This work was supported by the National
Key Research and Development Program of China through grants
2017YFA0402702 (C. M. and R. d. G.) and 2016YFA0400702 (L. C. H). We
also acknowledge research support from the National Natural Science
Foundation of China through grants 11373010 and U1631102 (C. M. and
R. d. G.), and 11303008 and 11473002 (L. C. H.).

\appendix

In this Appendix we provide our star cluster data in tabulated
form. For each galaxy, we provide one table containing the cluster
photometry and a second including the derived physical
parameters. Note that we do not provide the photometry for NGC 1512
and NGC 6951, given that those tables were already published by de
Grijs et al. (2017).

All tables are published in their entirety in the electronic version
of {\it The Astrophysical Journal}. Smaller portions are shown here
for guidance regarding their form and content.

\begin{table*}
\begin{center}
\begin{minipage}{140mm}
\caption{Multi-band photometry of the ESO 565-11 ring cluster candidates} 
{\tiny
\begin{tabular}{cccccccccc}
\hline
\#&\multicolumn{2}{c}{R.A. (J2000)} & \multicolumn{2}{c}{Dec. (J2000)}&$m_{\rm F255W}$&$m_{\rm F336W}$&$m_{\rm F439W}$&$m_{\rm F555W}$&$m_{\rm F814W}$\\
 &$(^{\circ})$&(hh mm ss.ss)&$(^{\circ})$& $(^{\circ} \, ' \, '')$ & (mag) & (mag) & (mag) & (mag)&(mag)\\
\hline
  0 &142.3189& 9 29 16.54& $-$20.3805& $-$20 22 49.98& 23.92$\pm$4.11& ------& 25.00$\pm$1.44& 26.58$\pm$1.23& ------\\
  1 &142.3178& 9 29 16.29& $-$20.3805& $-$20 22 50.03& 21.60$\pm$1.32& 22.15$\pm$0.51& 22.49$\pm$0.44& 22.94$\pm$0.18& 24.10$\pm$0.27\\
  2& 142.3193& 9 29 16.64& $-$20.3805& $-$20 22 49.81& 21.47$\pm$1.23& 23.79$\pm$1.12& 24.55$\pm$1.18& 24.56$\pm$0.43& 26.51$\pm$1.04\\
  3 &142.3178& 9 29 16.29& $-$20.3804& $-$20 22 49.80& 22.92$\pm$2.49& 23.62$\pm$1.02& 23.99$\pm$0.88& 24.40$\pm$0.37& 25.48$\pm$0.52\\
  4& 142.3173& 9 29 16.15& $-$20.3804& $-$20 22 49.75& 21.43$\pm$1.24& 23.44$\pm$0.97& 23.84$\pm$0.84& 23.93$\pm$0.32& 24.76$\pm$0.40\\
 $\cdots$ & $\cdots$ & $\cdots$ & $\cdots$ & $\cdots$ & $\cdots$ & $\cdots$ & $\cdots$ & $\cdots$&$\cdots$ \\
\hline
\end{tabular}
}
\end{minipage}
\end{center}
\end{table*}

\begin{table}
\begin{center}
\caption{Derived ESO 565-11 ring star cluster properties}
\begin{tabular}{cccccccccc}
\hline
\#&\multicolumn{3}{c}{log($t$ $\rm{yr}^{-1}$)}& &\multicolumn{3}{c}{log($M_{\rm cl} / M_{\odot}$)} & &$E(B-V)$\\
\cline{2-4} \cline{6-8}
\colhead{}&\colhead{Best}&\colhead{Min.}&\colhead{Max.}&\colhead{}&\colhead{Best}&\colhead{Min.}&\colhead{Max.}&\colhead{}&\colhead{(mag)}\\
\hline
0 &6.60&6.60& 10.18&& 6.00& 6.00& 6.00&& 0.80\\
1& 6.60&6.60 &8.16&& 5.00& 4.63& 6.04&& 0.20\\
2& 6.60& 6.60& 10.18&& 5.34& 4.67& 10.20 &&0.80\\
3 &6.60& 6.60&  9.81&& 4.58& 4.10& 6.83&& 0.30\\
4& 6.60& 6.60&  8.73&& 4.93& 4.38& 6.11&& 0.40\\
$\cdots$ & $\cdots$ & $\cdots$ & $\cdots$ && $\cdots$ & $\cdots$ & $\cdots$ & & $\cdots$ \\
\hline
\end{tabular}
\end{center}
\end{table}

\begin{table*}
\begin{center}
\begin{minipage}{140mm}
\caption{Multi-band photometry of the IC 342 ring cluster candidates} 
{\tiny
\begin{tabular}{cccccccccc}
\hline
\#&\multicolumn{2}{c}{R.A. (J2000)} & \multicolumn{2}{c}{Dec. (J2000)}&$m_{\rm F275W}$&$m_{\rm F336W}$&$m_{\rm F438W}$&$m_{\rm F547M}$&$m_{\rm F814W}$\\
 &$(^{\circ})$&(hh mm ss.ss)&$(^{\circ})$& $(^{\circ} \, ' \, '')$ & (mag) & (mag) & (mag) & (mag)&(mag)\\
\hline
  0& 56.7021& 3 46 48.52& 68.0955& 68  5  43.80& 22.26$\pm$0.74& 21.68$\pm$0.37& 21.37$\pm$0.23& 21.38$\pm$0.21& 21.72$\pm$0.18\\
  1& 56.7019& 3 46 48.46& 68.0956& 68  5  44.39& 21.74$\pm$0.63& 21.62$\pm$0.50& 22.00$\pm$0.65& 22.08$\pm$0.62& 22.91$\pm$0.88\\
  2& 56.7021& 3 46 48.50& 68.0956& 68  5  44.43& 20.91$\pm$0.43& 20.81$\pm$0.33& 20.73$\pm$0.26& 20.45$\pm$0.18& 20.28$\pm$0.09\\
  3& 56.7009& 3 46 48.23& 68.0957& 68  5  44.54& 22.86$\pm$1.03& 22.58$\pm$0.67& 21.79$\pm$0.33& 21.47$\pm$0.24& 21.71$\pm$0.21\\
  4& 56.7021& 3 46 48.52& 68.0957&68  5  44.56& 19.78$\pm$0.29&19.23$\pm$0.18& 19.18$\pm$0.16& 19.25$\pm$0.16& 19.39$\pm$0.12\\
$\cdots$ & $\cdots$ & $\cdots$ & $\cdots$ & $\cdots$ & $\cdots$ & $\cdots$ & $\cdots$ & $\cdots$&$\cdots$ \\
\hline
\end{tabular}
}
\end{minipage}
\end{center}
\tablecomments{}
\end{table*}

\begin{table}
\begin{center}
\caption{Derived IC 342 ring star cluster properties}
\begin{tabular}{cccccccccc}
\hline
\#&\multicolumn{3}{c}{log($t$ $\rm{yr}^{-1}$)}& &\multicolumn{3}{c}{log($M_{\rm cl} / M_{\odot}$)} & &$E(B-V)$\\
\cline{2-4} \cline{6-8}
\colhead{}&\colhead{Best}&\colhead{Min.}&\colhead{Max.}&\colhead{}&\colhead{Best}&\colhead{Min.}&\colhead{Max.}&\colhead{}&\colhead{(mag)}\\
\hline
0& 6.60& 6.60&  8.95&& 3.96& 3.40& 4.70&& 0.80\\
1&6.60& 6.60&  8.36&& 3.20& 2.43& 3.90 &&0.50\\
2& 6.90& 6.60&  8.20&& 4.03& 3.82& 5.20&& 0.50\\
3& 9.38& 6.60& 10.04&& 4.93& 3.32& 5.56&& 0.00\\
4 &6.60& 6.60&8.37&& 4.69& 4.13& 5.49&& 0.70\\
$\cdots$ & $\cdots$ & $\cdots$ & $\cdots$ && $\cdots$ & $\cdots$ & $\cdots$ & & $\cdots$ \\
\hline
\end{tabular}
\end{center}
\end{table}

\begin{table*}
\begin{center}
\begin{minipage}{130mm}
\caption{Multi-band photometry of the NGC 1097 ring cluster candidates} 
{\tiny
\begin{tabular}{ccccccccc}
\hline
\#&\multicolumn{2}{c}{R.A. (J2000)} & \multicolumn{2}{c}{Dec. (J2000)}&$m_{\rm F336W}$&$m_{\rm F438W}$&$m_{\rm F547M}$&$m_{\rm F814W}$\\
 &$(^{\circ})$&(hh mm ss.ss)&$(^{\circ})$& $(^{\circ} \, ' \, '')$ & (mag) & (mag) & (mag) & (mag)\\
\hline
  0 &41.5828& 2 46 19.87& $-$30.2772& $-$30 16 38.27 &23.46$\pm$0.83& 22.42$\pm$0.37& 22.85$\pm$0.38& 23.61$\pm$0.32\\
  1& 41.5812& 2 46 19.49& $-$30.2784& $-$30 16 42.29&------& 25.36$\pm$1.51& 24.14$\pm$0.69& 23.77$\pm$0.33\\  
  2& 41.5815& 2 46 19.56& $-$30.2779& $-$30 16 40.57& 24.10$\pm$1.13& 24.49$\pm$1.09& 23.54$\pm$0.54& 24.26$\pm$0.46\\  
  3& 41.5834& 2 46 20.02& $-$30.2763& $-$30 16 34.74& 25.24$\pm$1.91& 24.31$\pm$0.90& 24.28$\pm$0.73& 23.98$\pm$0.36\\
 4 &41.5829& 2 46 19.91& $-$30.2766& $-$30 16 35.89& 24.00$\pm$1.07& 23.39$\pm$0.58& 23.25$\pm$0.45& 23.58$\pm$0.30\\
 $\cdots$ & $\cdots$ & $\cdots$ & $\cdots$ & $\cdots$ & $\cdots$ & $\cdots$ & $\cdots$ & $\cdots$\\
\hline
\end{tabular}
}
\end{minipage}
\end{center}
\end{table*}

\begin{table}
\begin{center}
\caption{Derived NGC 1097 ring star cluster properties}
\begin{tabular}{cccccccccc}
\hline
\#&\multicolumn{3}{c}{log($t$ $\rm{yr}^{-1}$)}& &\multicolumn{3}{c}{log($M_{\rm cl} / M_{\odot}$)} & &$E(B-V)$\\
\cline{2-4} \cline{6-8}
\colhead{}&\colhead{Best}&\colhead{Min.}&\colhead{Max.}&\colhead{}&\colhead{Best}&\colhead{Min.}&\colhead{Max.}&\colhead{}&\colhead{(mag)}\\
\hline
  0 &8.58& 6.60 & 9.05&& 5.16& 4.00& 5.67&& 0.00\\
  1& 6.60& 6.60& 10.18&& 4.91& 4.25& 8.75 &&1.10\\
  2& 6.60& 6.60& 10.18&& 4.00& 3.47& 6.39&& 0.40\\
  3 &7.45& 6.60& 10.18&& 4.87& 3.89& 6.43&& 0.60\\
  4& 9.00& 6.60& 10.18&& 5.50& 3.73& 6.43&& 0.10\\
$\cdots$ & $\cdots$ & $\cdots$ & $\cdots$ && $\cdots$ & $\cdots$ & $\cdots$ & & $\cdots$ \\
\hline
\end{tabular}
\end{center}
\end{table}

\begin{table*}
\begin{center}
\begin{minipage}{140mm}
\caption{Multi-band photometry of the NGC 1326 ring cluster candidates} 
{\tiny
\begin{tabular}{cccccccccc}
\hline
\#&\multicolumn{2}{c}{R.A. (J2000)} & \multicolumn{2}{c}{Dec. (J2000)}&$m_{\rm F255W}$&$m_{\rm F336W}$&$m_{\rm F439W}$&$m_{\rm F555W}$&$m_{\rm F814W}$\\
 &$(^{\circ})$&(hh mm ss.ss)&$(^{\circ})$& $(^{\circ} \, ' \, '')$ & (mag) & (mag) & (mag) & (mag)&(mag)\\
\hline
 0& 50.9851& 3 23 56.42& $-$36.4654& $-$36 27 55.45& ------& ------& 23.44$\pm$0.89& 22.74$\pm$0.23& 23.65$\pm$0.38\\
1& 50.9845& 3 23 56.30& $-$36.4653& $-$36 27 55.33& 18.44$\pm$0.31& 19.04$\pm$0.14& 19.68$\pm$0.18& 20.46$\pm$0.24& 21.66$\pm$0.47\\
  2& 50.9857& 3 23 56.57& $-$36.4653& $-$36 27 55.23& 21.16$\pm$1.24& 20.43$\pm$0.25& 20.42$\pm$0.22& 20.55$\pm$0.15& 20.99$\pm$0.16\\
 3& 50.9848& 3 23 56.37& $-$36.4653& $-$36 27 55.25& 21.82$\pm$1.63& 21.62$\pm$0.45& 21.27$\pm$0.30& 21.72$\pm$0.22& 22.51$\pm$0.31\\
 4& 50.9857& 3 23 56.57& $-$36.4653& $-$36 27 55.23& 21.81$\pm$1.78& 20.89$\pm$0.31& 20.91$\pm$0.27& 21.27$\pm$0.19& 21.48$\pm$0.19\\ 

 $\cdots$ & $\cdots$ & $\cdots$ & $\cdots$ & $\cdots$ & $\cdots$ & $\cdots$ & $\cdots$ & $\cdots$&$\cdots$\\
\hline
\end{tabular}
}
\end{minipage}
\end{center}
\end{table*}

\begin{table}
\begin{center}
\caption{Derived NGC 1326 ring star cluster properties}
\begin{tabular}{cccccccccc}
\hline
\#&\multicolumn{3}{c}{log($t$ $\rm{yr}^{-1}$)}& &\multicolumn{3}{c}{log($M_{\rm cl} / M_{\odot}$)} & &$E(B-V)$\\
\cline{2-4} \cline{6-8}
\colhead{}&\colhead{Best}&\colhead{Min.}&\colhead{Max.}&\colhead{}&\colhead{Best}&\colhead{Min.}&\colhead{Max.}&\colhead{}&\colhead{(mag)}\\
\hline
 0 &6.60& 6.60& 10.18&& 4.41& 3.43& 6.30&& 0.70\\
 1& 6.60& 6.60&  6.60&& 4.35& 4.35& 4.35&& 0.00\\
 2 &6.90& 6.60&  8.46&& 4.80& 4.43& 6.11&& 0.20\\
 3& 8.15& 6.60&  8.65&& 5.14& 3.96& 5.57&& 0.00\\
 4& 7.20& 6.60&  8.38&& 5.02& 4.22& 5.87&& 0.20\\
$\cdots$ & $\cdots$ & $\cdots$ & $\cdots$ && $\cdots$ & $\cdots$ & $\cdots$ & & $\cdots$ \\
\hline
\end{tabular}
\end{center}
\end{table}

\begin{table*}
\begin{center}
\begin{minipage}{140mm}
\caption{Multi-band photometry of the NGC 1672 ring cluster candidates} 
{\tiny
\begin{tabular}{cccccccccc}
\hline
\#&\multicolumn{2}{c}{R.A. (J2000)} & \multicolumn{2}{c}{Dec. (J2000)}&$m_{\rm F330W}$&$m_{\rm F435W}$&$m_{\rm F550M}$&$m_{\rm F606W}$&$m_{\rm F814W}$\\
 &$(^{\circ})$&(hh mm ss.ss)&$(^{\circ})$& $(^{\circ} \, ' \, '')$ & (mag) & (mag) & (mag) & (mag)&(mag)\\
\hline
0& 71.4280& 4 45 42.73& $-$59.2485& $-$59 14 54.68& 19.82$\pm$0.17&19.69$\pm$0.08& 19.94$\pm$0.11& 19.93$\pm$0.04& 20.52$\pm$0.09\\
  1& 71.4274& 4 45 42.59& $-$59.2485& $-$59 14 54.66& 19.08$\pm$0.12& 19.52$\pm$0.08& 20.21$\pm$0.13& 20.33$\pm$0.06& 21.48$\pm$0.12\\
2& 71.4281& 4 45 42.74& $-$59.2484& $-$59 14 54.48& 19.90$\pm$0.18& 19.78$\pm$0.10& 20.15$\pm$0.13& 20.18$\pm$0.07& 20.76$\pm$0.09\\
   3& 71.4294& 4 45 43.06& $-$49.2484& $-$59 14 54.40& 23.14$\pm$0.83& 22.84$\pm$0.42& 22.48$\pm$0.39& 22.26$\pm$0.17& 22.36$\pm$0.18\\
  4 &71.4284& 4 45 42.82& $-$59.2484& $-$59 14 54.37& 20.90$\pm$0.29& 20.54$\pm$0.15& 21.14$\pm$0.25& 21.24$\pm$0.18& 22.25$\pm$0.46\\
 $\cdots$ & $\cdots$ & $\cdots$ & $\cdots$ & $\cdots$ & $\cdots$ & $\cdots$ & $\cdots$ & $\cdots$&$\cdots$\\
\hline
\end{tabular}
}
\end{minipage}
\end{center}
\end{table*}

\begin{table}
\begin{center}
\caption{Derived NGC 1672 ring star cluster properties}
\begin{tabular}{cccccccccc}
\hline
\#&\multicolumn{3}{c}{log($t$ $\rm{yr}^{-1}$)}& &\multicolumn{3}{c}{log($M_{\rm cl} / M_{\odot}$)} & &$E(B-V)$\\
\cline{2-4} \cline{6-8}
\colhead{}&\colhead{Best}&\colhead{Min.}&\colhead{Max.}&\colhead{}&\colhead{Best}&\colhead{Min.}&\colhead{Max.}&\colhead{}&\colhead{(mag)}\\
\hline
0& 6.60& 6.60&  8.49&& 5.37& 4.86& 6.23&& 0.50\\
1 &6.60& 6.60&  6.60&& 4.61& 4.61& 4.79&& 0.10\\
2& 7.78& 6.60 & 8.40&& 5.70& 4.62& 6.04&& 0.10\\
3& 6.60 &6.60 &10.05&&5.18& 4.33& 6.54&& 1.00\\
4& 8.11& 6.60&  8.51&& 5.34& 4.19& 5.64&& 0.00\\

$\cdots$ & $\cdots$ & $\cdots$ & $\cdots$ && $\cdots$ & $\cdots$ & $\cdots$ & & $\cdots$ \\
\hline
\end{tabular}
\end{center}
\end{table}

\begin{table*}
\begin{center}
\begin{minipage}{160mm}
\caption{Multi-band photometry of the NGC 2997 ring cluster candidates} 
{\tiny
\begin{tabular}{ccccccccccc}
\hline
\#&\multicolumn{2}{c}{R.A. (J2000)} & \multicolumn{2}{c}{Dec. (J2000)}&$m_{\rm F220W}$&$m_{\rm F330W}$&$m_{\rm F336W}$&$m_{\rm F555W}$&$m_{\rm F606W}$&$m_{\rm F814W}$\\
 &$(^{\circ})$&(hh mm ss.ss)&$(^{\circ})$& $(^{\circ} \, ' \, '')$ & (mag) & (mag) & (mag) & (mag)&(mag)&(mag)\\
\hline
0 &146.4102& 9 45 38.44& $-$31.1929& $-$31 11 34.50& 25.72$\pm$5.18& 25.42$\pm$2.31& 25.34$\pm$1.98& 25.58$\pm$0.95& 26.21$\pm$0.71& 28.24$\pm$2.55\\
  1& 146.4110& 9 45 38.65& $-$31.1928& $-$31 11 34.25& ------& 25.76$\pm$2.71& 25.96$\pm$2.63& 24.27$\pm$0.50& 25.66$\pm$0.50& 25.25$\pm$0.52\\
  2 &146.4107 &9 45 38.59& $-$31.1926& $-$31 11 33.40& ------& 24.21$\pm$1.32& 24.25$\pm$1.20& 24.98$\pm$0.72& 25.51$\pm$0.48& 25.78$\pm$0.68\\
  3& 146.4105& 9 45 38.52& $-$31.1925& $-$31 11 33.31& 22.51$\pm$1.15& 22.71$\pm$0.66& 22.96$\pm$0.66& 23.88$\pm$0.42& 24.29$\pm$0.26& 25.39$\pm$0.56\\
  4& 146.4104& 9 45 38.50& $-$31.1925& $-$31 11 33.32& 22.56$\pm$1.18& 22.67$\pm$0.65& 23.04$\pm$0.69& 24.05$\pm$0.48& 25.22$\pm$0.53& 25.59$\pm$0.69\\
$\cdots$ & $\cdots$ & $\cdots$ & $\cdots$ & $\cdots$ & $\cdots$ & $\cdots$ & $\cdots$ & $\cdots$&$\cdots$&$\cdots$\\
\hline
\end{tabular}
}
\end{minipage}
\end{center}
\end{table*}

\begin{table}
\begin{center}
\caption{Derived NGC 2997 ring star cluster properties}
\begin{tabular}{cccccccccc}
\hline
\#&\multicolumn{3}{c}{log($t$ $\rm{yr}^{-1}$)}& &\multicolumn{3}{c}{log($M_{\rm cl} / M_{\odot}$)} & &$E(B-V)$\\
\cline{2-4} \cline{6-8}
\colhead{}&\colhead{Best}&\colhead{Min.}&\colhead{Max.}&\colhead{}&\colhead{Best}&\colhead{Min.}&\colhead{Max.}&\colhead{}&\colhead{(mag)}\\
\hline
0 &6.60 & 6.60 & 10.18 & & 1.89 & 1.89&7.95 && 0.00\\
1 & 7.20 & 6.60 & 10.18&& 3.00& 2.52& 5.39& &0.00\\
2 &8.00& 6.60& 10.18&& 3.43& 2.27& 5.38&& 0.00\\
3 &6.60& 6.60&  7.78&& 2.62& 2.62& 3.80&& 0.00\\
4&6.60& 6.60&  7.60&& 2.51& 2.51& 3.55&& 0.00\\

$\cdots$ & $\cdots$ & $\cdots$ & $\cdots$ && $\cdots$ & $\cdots$ & $\cdots$ & & $\cdots$ \\
\hline
\end{tabular}
\end{center}
\end{table}

\begin{table*}
\begin{center}
\begin{minipage}{140mm}
\caption{Multi-band photometry of the NGC 3081 ring cluster candidates} 
{\tiny
\begin{tabular}{cccccccccc}
\hline
\#&\multicolumn{2}{c}{R.A. (J2000)} & \multicolumn{2}{c}{Dec. (J2000)}&$m_{\rm F255W}$&$m_{\rm F336W}$&$m_{\rm F439W}$&$m_{\rm F555W}$&$m_{\rm F814W}$\\
 &$(^{\circ})$&(hh mm ss.ss)&$(^{\circ})$& $(^{\circ} \, ' \, '')$ & (mag) & (mag) & (mag) & (mag)&(mag)\\
\hline
0 &149.8739& 9 59 29.74& $-$22.8274& $-$22 49 38.98& 20.23$\pm$0.68& 20.83$\pm$0.28& 21.38$\pm$0.26& 22.12$\pm$0.14& 23.23$\pm$0.24\\
1& 149.8736 &9 59 29.66& $-$22.82746& $-$22 49 38.88& 21.52$\pm$1.26& 21.49$\pm$0.38& 21.51$\pm$0.28& 21.92$\pm$0.17& 22.19$\pm$0.22\\
2 &149.8744& 9 59 29.86& $-$22.8273& $-$22 49 38.62 &22.37$\pm$2.23& 21.07$\pm$0.35&19.97$\pm$0.18& 19.82$\pm$0.13& 19.70$\pm$0.11\\
 3 &149.8733 &9 59 29.60& $-$22.8273& $-$22 49 38.39& 20.79$\pm$0.89& 20.99$\pm$0.31& 21.25$\pm$0.26& 21.45$\pm$0.14& 21.89$\pm$0.21\\
 4& 149.8724 &9 59 29.39& $-$22.8272& $-$22 49 38.14& 22.93$\pm$2.42& 24.64$\pm$1.81& 23.80$\pm$0.88& 24.26$\pm$0.63& 26.90$\pm$5.20\\
$\cdots$ & $\cdots$ & $\cdots$ & $\cdots$ & $\cdots$ & $\cdots$ & $\cdots$ & $\cdots$ & $\cdots$&$\cdots$\\
\hline
\end{tabular}
}
\end{minipage}
\end{center}
\end{table*}

\begin{table}
\begin{center}
\caption{Derived NGC 3081 ring star cluster properties}
\begin{tabular}{cccccccccc}
\hline
\#&\multicolumn{3}{c}{log($t$ $\rm{yr}^{-1}$)}& &\multicolumn{3}{c}{log($M_{\rm cl} / M_{\odot}$)} & &$E(B-V)$\\
\cline{2-4} \cline{6-8}
\colhead{}&\colhead{Best}&\colhead{Min.}&\colhead{Max.}&\colhead{}&\colhead{Best}&\colhead{Min.}&\colhead{Max.}&\colhead{}&\colhead{(mag)}\\
\hline
0 &6.60& 6.60&  7.30&& 4.37& 4.17& 4.99&& 0.10\\
1 &7.30 &6.60& 8.40 &&5.38 &4.46 &6.04 &&0.20\\
2& 8.13& 6.60& 10.07&& 7.37& 6.71& 8.10&& 0.60\\
3& 7.30& 6.60&  8.16&& 5.52& 4.76& 5.97&& 0.20\\
4&6.60& 6.60& 10.18&& 3.61& 3.23& 10.11&& 0.20\\
$\cdots$ & $\cdots$ & $\cdots$ & $\cdots$ && $\cdots$ & $\cdots$ & $\cdots$ & & $\cdots$ \\
\hline
\end{tabular}
\end{center}
\end{table}

\begin{table*}
\begin{center}
\begin{minipage}{160mm}
\caption{Multi-band photometry of the NGC 3351 ring cluster candidates} 
{\tiny
\begin{tabular}{cccccccccccc}
\hline
\#&\multicolumn{2}{c}{R.A. (J2000)} & \multicolumn{2}{c}{Dec. (J2000)}&$m_{\rm F275W}$&$m_{\rm F336W}$&$m_{\rm F438W}$&$m_{\rm F450W}$&$m_{\rm F555W}$&$m_{\rm F606W}$&$m_{\rm F814W}$\\
 &$(^{\circ})$&(hh mm ss.ss)&$(^{\circ})$& $(^{\circ} \, ' \, '')$ & (mag) & (mag) & (mag) & (mag)&(mag)&(mag)&(mag)\\
\hline
0 &160.9910& 10 43 57.85& 11.7014& 11 42   5.28& 23.02$\pm$1.064& 22.54$\pm$0.55& 21.79$\pm$0.27& 21.96$\pm$0.19& 21.89$\pm$0.15& 22.09$\pm$0.20& 22.49$\pm$0.18\\
  1 &160.9907 &10 43 57.77 &11.7014& 11 42   5.30 &22.48$\pm$0.82& 23.21$\pm$0.75& 23.91$\pm$0.83& 24.08$\pm$0.66& 24.57$\pm$0.82&25.03$\pm$1.15&------\\
  2 &160.9905& 10 43 57.72 &11.7015& 11 42   5.64 &21.96$\pm$0.65& 21.88$\pm$0.40& 22.05$\pm$0.31 &22.31$\pm$0.24& 22.34$\pm$0.18& 22.77$\pm$0.27& 23.74$\pm$0.41\\
  3& 160.9909& 10 43 57.83& 11.7016& 11 42   5.89& 22.41$\pm$0.80& 23.04$\pm$0.71&23.18$\pm$0.57&23.76$\pm$0.53&23.95$\pm$0.52& 24.52$\pm$0.74&25.39$\pm$0.95\\
  4& 160.9903 &10 43 57.69 &11.7016& 11 42   5.93& 20.00$\pm$0.26& 20.24$\pm$0.19&20.04$\pm$0.12&20.29$\pm$0.09& 20.19$\pm$0.07& 20.35$\pm$0.09&20.70$\pm$0.08\\
$\cdots$ & $\cdots$ & $\cdots$ & $\cdots$ & $\cdots$ & $\cdots$ & $\cdots$ & $\cdots$ & $\cdots$&$\cdots$&$\cdots$&$\cdots$\\
\hline
\end{tabular}
}
\end{minipage}
\end{center}
\end{table*}

\begin{table}
\begin{center}
\caption{Derived NGC 3351 ring star cluster properties}
\begin{tabular}{cccccccccc}
\hline
\#&\multicolumn{3}{c}{log($t$ $\rm{yr}^{-1}$)}& &\multicolumn{3}{c}{log($M_{\rm cl} / M_{\odot}$)} & &$E(B-V)$\\
\cline{2-4} \cline{6-8}
\colhead{}&\colhead{Best}&\colhead{Min.}&\colhead{Max.}&\colhead{}&\colhead{Best}&\colhead{Min.}&\colhead{Max.}&\colhead{}&\colhead{(mag)}\\
\hline
0 &8.89 &6.60&  9.13&& 5.27& 3.84& 5.53& &0.00\\
1&6.60& 6.60&  7.88&& 2.32& 2.32& 3.74&& 0.00\\
2 &6.60& 6.60&  8.19&&3.60& 3.29& 4.53&& 0.30\\
3& 6.60& 6.60&  7.90&& 2.68& 2.48& 3.78&& 0.10\\
4& 6.90& 6.90&  8.09&& 4.55& 4.55& 5.71&& 0.20\\
$\cdots$ & $\cdots$ & $\cdots$ & $\cdots$ && $\cdots$ & $\cdots$ & $\cdots$ & & $\cdots$ \\
\hline
\end{tabular}
\end{center}
\end{table}

\begin{table*}
\begin{center}
\begin{minipage}{110mm}
\caption{Multi-band photometry of the NGC 4303 ring cluster candidates} 
{\tiny
\begin{tabular}{cccccccc}
\hline
\#&\multicolumn{2}{c}{R.A. (J2000)} & \multicolumn{2}{c}{Dec. (J2000)}&$m_{\rm F330W}$&$m_{\rm F555W}$&$m_{\rm F814W}$\\
 &$(^{\circ})$&(hh mm ss.ss)&$(^{\circ})$& $(^{\circ} \, ' \, '')$ & (mag) & (mag) & (mag)\\
\hline
  0 &185.4785& 12 21 54.85  &4.4727&  4 28  21.86 &23.88$\pm$1.13& 23.58$\pm$0.39& 23.66$\pm$0.26\\
  1 &185.4792& 12 21 55.02&  4.4727&  4 28  21.94& 24.20$\pm$1.31& 25.01$\pm$0.93& 25.32$\pm$0.64\\
  2& 185.4789& 12 21 54.95&  4.4728&  4 28  22.08& 22.89$\pm$0.71& 23.18$\pm$0.30& 24.19$\pm$0.34\\
  3 &185.4790& 12 21 54.98&  4.4729&  4 28  22.52& 21.89$\pm$0.45& 21.66$\pm$0.19& 23.04$\pm$0.28\\
  4& 185.4793 &12 21 55.04&  4.4729 & 4 28  22.53& 22.27$\pm$0.54& 22.47$\pm$0.22& 23.63$\pm$0.25\\

$\cdots$ & $\cdots$ & $\cdots$ & $\cdots$ & $\cdots$ & $\cdots$ & $\cdots$ & $\cdots$\\
\hline
\end{tabular}
}
\end{minipage}
\end{center}
\end{table*}

\begin{table}
\begin{center}
\caption{Derived NGC 4303 ring star cluster properties}
\begin{tabular}{cccccccccc}
\hline
\#&\multicolumn{3}{c}{log($t$ $\rm{yr}^{-1}$)}& &\multicolumn{3}{c}{log($M_{\rm cl} / M_{\odot}$)} & &$E(B-V)$\\
\cline{2-4} \cline{6-8}
\colhead{}&\colhead{Best}&\colhead{Min.}&\colhead{Max.}&\colhead{}&\colhead{Best}&\colhead{Min.}&\colhead{Max.}&\colhead{}&\colhead{(mag)}\\
\hline

0& 6.60 &6.60& 10.18&& 4.56& 3.54& 6.38&& 0.90\\
1& 6.60& 6.60& 10.18&& 3.81& 2.79& 8.10&& 0.80\\
2& 6.60& 6.60&  8.59&& 3.86& 3.33& 4.96&& 0.40\\
3& 6.60& 6.60&  8.68&& 4.35& 3.85& 5.52&& 0.40\\
4&6.60& 6.60&  8.49&& 4.12& 3.62& 5.11&& 0.40\\
$\cdots$ & $\cdots$ & $\cdots$ & $\cdots$ && $\cdots$ & $\cdots$ & $\cdots$ & & $\cdots$ \\
\hline
\end{tabular}
\end{center}
\end{table}

\clearpage

\begin{table*}
\begin{center}
\begin{minipage}{140mm}
\caption{Multi-band photometry of the NGC 4314 ring cluster candidates} 
{\tiny
\begin{tabular}{cccccccccc}
\hline
\#&\multicolumn{2}{c}{R.A. (J2000)} & \multicolumn{2}{c}{Dec. (J2000)}&$m_{\rm F336W}$&$m_{\rm F439W}$&$m_{\rm F569W}$&$m_{\rm F606W}$&$m_{\rm F814W}$\\
 &$(^{\circ})$&(hh mm ss.ss)&$(^{\circ})$& $(^{\circ} \, ' \, '')$ & (mag) & (mag) & (mag) & (mag)&(mag)\\
\hline
0& 185.6327& 12 22 31.86& 29.8932& 29 53  35.77 &23.28$\pm$0.87& 23.45$\pm$0.69&24.39$\pm$0.42&24.49$\pm$0.30& 25.02$\pm$0.48\\
1& 185.6340& 12 22 32.16& 29.8934& 29 53  36.40& 22.22$\pm$0.55&21.23$\pm$0.25&21.90$\pm$0.13& 22.24$\pm$0.11& 22.43$\pm$0.13\\
 2 & 185.6345 &12 22 32.28& 29.8935& 29 53  36.65& 22.08$\pm$0.51& 22.17$\pm$0.39& 23.08$\pm$0.27& 23.72$\pm$0.31& 23.56$\pm$0.28\\
3& 185.6342& 12 22 32.21& 29.8935& 29 53  36.66&21.64$\pm$0.41& 21.63$\pm$0.30& 22.19$\pm$0.15& 22.41$\pm$0.11& 22.58$\pm$0.14\\
4& 185.6344& 12 22 32.26 &29.8935& 29 53  36.93& 22.00$\pm$0.51& 22.28$\pm$0.44& 23.33$\pm$0.45& 23.45$\pm$0.36& 22.94$\pm$0.21\\
$\cdots$ & $\cdots$ & $\cdots$ & $\cdots$ & $\cdots$ & $\cdots$ & $\cdots$ & $\cdots$ & $\cdots$&$\cdots$\\
\hline
\end{tabular}
}
\end{minipage}
\end{center}
\end{table*}

\clearpage

\begin{table}
\begin{center}
\caption{Derived NGC 4314 ring star cluster properties}
\begin{tabular}{cccccccccc}
\hline
\#&\multicolumn{3}{c}{log($t$ $\rm{yr}^{-1}$)}& &\multicolumn{3}{c}{log($M_{\rm cl} / M_{\odot}$)} & &$E(B-V)$\\
\cline{2-4} \cline{6-8}
\colhead{}&\colhead{Best}&\colhead{Min.}&\colhead{Max.}&\colhead{}&\colhead{Best}&\colhead{Min.}&\colhead{Max.}&\colhead{}&\colhead{(mag)}\\
\hline
0&7.30& 6.60&  8.38&& 3.25& 2.47& 4.03&& 0.00\\
1&8.30& 7.56&  8.75&& 5.00& 4.58& 5.21&& 0.10\\
2&7.30& 6.60&  8.19&& 3.73& 3.16& 4.36&& 0.00\\
3& 7.38& 6.90&  8.31&& 4.28& 3.49& 4.84&& 0.10\\
4& 7.20& 6.90&  8.20&& 3.83& 3.21& 4.62&& 0.10\\

$\cdots$ & $\cdots$ & $\cdots$ & $\cdots$ && $\cdots$ & $\cdots$ & $\cdots$ & & $\cdots$ \\
\hline
\end{tabular}
\end{center}
\end{table}

\begin{table*}
\begin{center}
\begin{minipage}{140mm}
\caption{Multi-band photometry of the NGC 6782 ring cluster candidates} 
{\tiny
\begin{tabular}{cccccccccc}
\hline
\#&\multicolumn{2}{c}{R.A. (J2000)} & \multicolumn{2}{c}{Dec. (J2000)}&$m_{\rm F255W}$&$m_{\rm F300W}$&$m_{\rm F450W}$&$m_{\rm F606W}$&$m_{\rm F814W}$\\
 &$(^{\circ})$&(hh mm ss.ss)&$(^{\circ})$& $(^{\circ} \, ' \, '')$ & (mag) & (mag) & (mag) & (mag)&(mag)\\
\hline

0& 290.9916& 19 23 57.99& $-$59.9239& $-$59 55 26.16&------& 23.21$\pm$0.90& 23.27$\pm$0.41& 23.73$\pm$0.33& 24.35$\pm$0.49\\
1& 290.9910& 19 23 57.84& $-$59.9238& $-$59 55 25.68 &21.01$\pm$0.99& 21.32$\pm$0.38& 22.31$\pm$0.35& 23.19$\pm$0.53& 24.55$\pm$1.41\\
2& 290.9905& 19 23 57.72& $-$59.9237& $-$59 55 25.49& 20.27$\pm$0.70& 20.89$\pm$0.30& 21.72$\pm$0.21& 22.14$\pm$0.19& 23.01$\pm$0.29\\
3 &290.9902& 19 23 57.65& $-$59.9237& $-$59 55 25.57& 19.97$\pm$0.62& 19.88$\pm$0.19& 20.23$\pm$0.12& 20.70$\pm$0.12& 21.26$\pm$0.15\\
4& 290.9928& 19 23 58.29& $-$59.9236& $-$59 55 25.23& 20.88$\pm$0.95& 21.18$\pm$0.35& 21.24$\pm$0.14& 21.49$\pm$0.09& 21.85$\pm$0.12\\
$\cdots$ & $\cdots$ & $\cdots$ & $\cdots$ & $\cdots$ & $\cdots$ & $\cdots$ & $\cdots$ & $\cdots$&$\cdots$\\
\hline
\end{tabular}
}
\end{minipage}
\end{center}
\end{table*}

\begin{table}
\caption{Derived NGC 6782 ring star cluster properties}
\begin{center}
\begin{tabular}{cccccccccc}
\hline
\#&\multicolumn{3}{c}{log($t$ $\rm{yr}^{-1}$)}& &\multicolumn{3}{c}{log($M_{\rm cl} / M_{\odot}$)} & &$E(B-V)$\\
\cline{2-4} \cline{6-8}
\colhead{}&\colhead{Best}&\colhead{Min.}&\colhead{Max.}&\colhead{}&\colhead{Best}&\colhead{Min.}&\colhead{Max.}&\colhead{}&\colhead{(mag)}\\
\hline
0 &8.11 &6.60&  8.58&& 5.49& 4.30& 5.94 &&0.00\\
1& 6.60& 6.60&  7.81&& 4.40& 4.40& 5.76&& 0.00\\
  2& 7.30& 6.60&  7.68&& 5.58& 4.95& 5.88&& 0.00\\
  3& 7.94& 6.60  &8.16&& 6.59& 5.71& 6.78&& 0.00\\
  4& 7.64& 6.60 & 8.30&& 6.38& 5.55& 6.65 &&0.20\\
$\cdots$ & $\cdots$ & $\cdots$ & $\cdots$ && $\cdots$ & $\cdots$ & $\cdots$ & & $\cdots$ \\
\hline
\end{tabular}
\end{center}
\end{table}

%
%

\begin{table}
\caption{Derived NGC 6951 ring star cluster properties}
\begin{center}
\begin{tabular}{cccccccccc}
\hline
\#&\multicolumn{3}{c}{log($t$ $\rm{yr}^{-1}$)}& &\multicolumn{3}{c}{log($M_{\rm cl} / M_{\odot}$)} & &$E(B-V)$\\
\cline{2-4} \cline{6-8}
\colhead{}&\colhead{Best}&\colhead{Min.}&\colhead{Max.}&\colhead{}&\colhead{Best}&\colhead{Min.}&\colhead{Max.}&\colhead{}&\colhead{(mag)}\\
\hline
0& 9.90& 6.60& 10.10&& 6.53 &4.93& 6.82&& 0.10\\
 1& 7.15& 6.60&  9.40&& 5.02&4.29& 5.87&& 0.60\\
 2& 7.25& 6.95&  7.35&& 6.15& 5.66& 6.36&& 0.80\\
 3&7.25& 6.90&  8.40&& 5.00& 4.37& 5.73&& 0.60\\
 4& 7.30& 6.95&  7.45&& 5.54& 4.99& 5.65&& 0.30\\
$\cdots$ & $\cdots$ & $\cdots$ & $\cdots$ && $\cdots$ & $\cdots$ & $\cdots$ & & $\cdots$ \\
\hline
\end{tabular}
\end{center}
\end{table}

\begin{table*}
\begin{center}
\begin{minipage}{140mm}
\caption{Multi-band photometry of the NGC 7217 ring cluster candidates} 
{\tiny
\begin{tabular}{ccccccccc}
\hline
\#&\multicolumn{2}{c}{R.A. (J2000)} & \multicolumn{2}{c}{Dec. (J2000)}&$m_{\rm F336W}$&$m_{\rm F450W}$&$m_{\rm F547M}$&$m_{\rm F814W}$\\
 &$(^{\circ})$&(hh mm ss.ss)&$(^{\circ})$& $(^{\circ} \, ' \, '')$ & (mag) & (mag) & (mag) & (mag)\\
\hline
 0 &331.9677& 22 7 52.26& 31.3564& 31 21  23.08 &24.75$\pm$1.73& 23.69$\pm$0.42& 24.03$\pm$0.47& 24.82$\pm$0.33\\
  1& 331.9665& 22 7 51.97& 31.3566& 31 21  23.82& 23.72$\pm$1.07& 23.86$\pm$0.49&24.13$\pm$0.55& 24.86$\pm$0.43\\
  2& 331.9663& 22 7 51.91& 31.3566& 31 21  23.95& 23.28$\pm$0.87& 23.01$\pm$0.31& 23.55$\pm$0.38& 24.94$\pm$0.37\\
3& 331.9690& 22 7 52.57& 31.3567& 31 21  24.46& 23.16$\pm$0.83& 23.03$\pm$0.33& 23.28$\pm$0.35& 23.77$\pm$0.23\\
  4& 331.9695& 22 7 52.69& 31.3568& 31 21  24.54&23.71$\pm$1.07& 24.21$\pm$0.57& 23.80$\pm$0.44& 25.24$\pm$0.46\\
$\cdots$ & $\cdots$ & $\cdots$ & $\cdots$ & $\cdots$ & $\cdots$ & $\cdots$ & $\cdots$ & $\cdots$\\
\hline
\end{tabular}
}
\end{minipage}
\end{center}
\end{table*}

\begin{table}
\caption{Derived NGC 7217 ring star cluster properties}
\begin{center}
\begin{tabular}{cccccccccc}
\hline
\#&\multicolumn{3}{c}{log($t$ $\rm{yr}^{-1}$)}& &\multicolumn{3}{c}{log($M_{\rm cl} / M_{\odot}$)} & &$E(B-V)$\\
\cline{2-4} \cline{6-8}
\colhead{}&\colhead{Best}&\colhead{Min.}&\colhead{Max.}&\colhead{}&\colhead{Best}&\colhead{Min.}&\colhead{Max.}&\colhead{}&\colhead{(mag)}\\
\hline
0& 8.45& 6.60&  9.72&& 4.64& 3.19& 5.76&& 0.00\\
1& 8.58& 6.60& 10.04&& 4.69& 3.15& 5.98&& 0.00\\ 
2 &6.60&6.60&  8.58&& 3.75& 3.22& 4.96&& 0.30\\ 
3& 7.64& 6.60&  8.94&& 4.74& 3.58& 5.44&& 0.20\\ 
4& 8.50& 6.60& 10.10&& 4.58& 3.09& 5.99&& 0.00\\ 
$\cdots$ & $\cdots$ & $\cdots$ & $\cdots$ && $\cdots$ & $\cdots$ & $\cdots$ & & $\cdots$ \\
\hline
\end{tabular}
\end{center}
\end{table}

\begin{table*}
\begin{center}
\begin{minipage}{140mm}
\caption{Multi-band photometry of the NGC 7469 ring cluster candidates} 
{\tiny
\begin{tabular}{cccccccccc}
\hline
\#&\multicolumn{2}{c}{R.A. (J2000)} & \multicolumn{2}{c}{Dec. (J2000)}&$m_{\rm F336W}$&$m_{\rm F435W}$&$m_{\rm F550M}$&$m_{\rm F606W}$&$m_{\rm F814W}$\\
 &$(^{\circ})$&(hh mm ss.ss)&$(^{\circ})$& $(^{\circ} \, ' \, '')$ & (mag) & (mag) & (mag) & (mag)&(mag)\\
\hline

 0& 345.8152& 23 3 15.64&  8.8737&  8 52  25.54& 18.46$\pm$0.09& 18.84$\pm$0.04&19.17$\pm$0.07&19.60$\pm$0.04& 20.40$\pm$0.05\\
  1& 345.8150& 23 3 15.62&  8.8737&  8 52  25.62& 19.64$\pm$0.16& 19.71$\pm$0.09&19.75$\pm$0.10& 20.00$\pm$0.07& 20.35$\pm$0.07\\
  2& 345.8152& 23 3 15.66&  8.8737&8 52  25.67& 20.22$\pm$0.22& 20.55$\pm$0.13& 20.31$\pm$0.13& 21.02$\pm$0.13& 21.56$\pm$ 0.14\\
  3& 345.8149& 23 3 15.58&  8.8737&  8 52  25.66 &19.96$\pm$0.18& 19.94$\pm$0.07& 20.28$\pm$0.11& 20.38$\pm$0.06& 20.91$\pm$0.08\\
  4& 345.8151& 23 3 15.63&  8.8738&  8 52  25.73& 21.15$\pm$0.32& 21.71$\pm$0.23& 22.05$\pm$0.33& 22.45$\pm$0.27& 23.26$\pm$0.51\\

$\cdots$ & $\cdots$ & $\cdots$ & $\cdots$ & $\cdots$ & $\cdots$ & $\cdots$ & $\cdots$ & $\cdots$&$\cdots$\\
\hline
\end{tabular}
}
\end{minipage}
\end{center}
\end{table*}

\begin{table}
\caption{Derived NGC 7469 ring star cluster properties}
\begin{center}
\begin{tabular}{cccccccccc}
\hline
\#&\multicolumn{3}{c}{log($t$ $\rm{yr}^{-1}$)}& &\multicolumn{3}{c}{log($M_{\rm cl} / M_{\odot}$)} & &$E(B-V)$\\
\cline{2-4} \cline{6-8}
\colhead{}&\colhead{Best}&\colhead{Min.}&\colhead{Max.}&\colhead{}&\colhead{Best}&\colhead{Min.}&\colhead{Max.}&\colhead{}&\colhead{(mag)}\\
\hline

0& 6.60 &6.60&  7.81&& 6.42& 6.42& 7.19&& 0.20\\
1& 6.90& 6.60&  8.27&& 6.37& 6.20& 7.52&& 0.20\\
2& 6.90& 6.60&  8.27&& 5.69& 5.69& 6.96&& 0.00\\
3& 7.72& 6.60&  8.28&& 6.89& 5.86& 7.21&& 0.10\\
4& 6.60& 6.60&  7.98&& 5.29& 4.94& 6.18&& 0.20\\
$\cdots$ & $\cdots$ & $\cdots$ & $\cdots$ && $\cdots$ & $\cdots$ & $\cdots$ & & $\cdots$ \\
\hline
\end{tabular}
\end{center}
\end{table}

\begin{table*}
\begin{center}
\begin{minipage}{140mm}
\caption{Multi-band photometry of the NGC 7742 ring cluster candidates} 
{\tiny
\begin{tabular}{ccccccccc}
\hline
\#&\multicolumn{2}{c}{R.A. (J2000)} & \multicolumn{2}{c}{Dec. (J2000)}&$m_{\rm F336W}$&$m_{\rm F555W}$&$m_{\rm F675W}$&$m_{\rm F814W}$\\
 &$(^{\circ})$&(hh mm ss.ss)&$(^{\circ})$& $(^{\circ} \, ' \, '')$ & (mag) & (mag) & (mag) & (mag)\\
\hline

0 &356.0654& 23 44 15.69& 10.7636& 10 45  49.27& 25.12$\pm$2.29& 23.94$\pm$0.42& 26.34$\pm$2.67& 25.16$\pm$0.75\\
  1& 356.0650& 23 44 15.62& 10.7637& 10 45  49.53& 23.97$\pm$1.21& 23.98$\pm$0.32& 25.62$\pm$1.05& 25.35$\pm$0.55\\
  2& 356.0667& 23 44 16.03& 10.7638& 10 45  49.83& 24.18$\pm$1.33& 24.93$\pm$0.61& 24.42$\pm$0.38& 25.86$\pm$0.83\\
  3&356.0663& 23 44 15.91& 10.7638& 10 45  49.97& 25.27$\pm$2.29& 25.40$\pm$0.68& 25.54$\pm$0.63& 26.04$\pm$0.72\\
  4& 356.0670& 23 44 16.09& 10.7638 &10 45  49.95&------& 24.52$\pm$0.40& 24.36$\pm$0.34& 24.15$\pm$0.27\\
$\cdots$ & $\cdots$ & $\cdots$ & $\cdots$ & $\cdots$ & $\cdots$ & $\cdots$ & $\cdots$ & $\cdots$\\
\hline
\end{tabular}
}
\end{minipage}
\end{center}
\end{table*}

\begin{table}
\caption{Derived NGC 7742 ring star cluster properties}
\begin{center}
\begin{tabular}{cccccccccc}
\hline
\#&\multicolumn{3}{c}{log($t$ $\rm{yr}^{-1}$)}& &\multicolumn{3}{c}{log($M_{\rm cl} / M_{\odot}$)} & &$E(B-V)$\\
\cline{2-4} \cline{6-8}
\colhead{}&\colhead{Best}&\colhead{Min.}&\colhead{Max.}&\colhead{}&\colhead{Best}&\colhead{Min.}&\colhead{Max.}&\colhead{}&\colhead{(mag)}\\
\hline
0& 6.60& 6.60& 10.18&& 4.06& 3.79& 6.67&& 0.20\\
1&6.60& 6.60& 10.18&& 3.88& 3.74& 6.50&&0.10\\
2& 6.60& 6.60& 10.18&& 4.23& 3.56& 8.89&& 0.50\\
3&6.60& 6.60& 10.18&& 4.13& 3.32& 8.65&& 0.60\\
4& 8.71& 6.90& 10.18&& 6.37& 4.71& 9.18&& 0.70\\

$\cdots$ & $\cdots$ & $\cdots$ & $\cdots$ && $\cdots$ & $\cdots$ & $\cdots$ & & $\cdots$ \\
\hline
\end{tabular}
\end{center}
\end{table}

\begin{table*}
\begin{center}
\begin{minipage}{120mm}
\caption{Multi-band photometry of the UGC 3789 ring cluster candidates} 
{\tiny
\begin{tabular}{cccccccc}
\hline
\#&\multicolumn{2}{c}{R.A. (J2000)} & \multicolumn{2}{c}{Dec. (J2000)}&$m_{\rm F336W}$&$m_{\rm F438W}$&$m_{\rm F814W}$\\
 &$(^{\circ})$&(hh mm ss.ss)&$(^{\circ})$& $(^{\circ} \, ' \, '')$ & (mag) & (mag) & (mag)\\
\hline
  0& 41.5856& 2 46 20.56& $-$30.2757& $-$30 16 32.72& 25.40$\pm$2.09& 24.62$\pm$1.06&24.71$\pm$0.53\\
  1& 41.5862& 2 46 20.68& $-$30.2752& $-$30 16 31.03& 23.89$\pm$1.04& 23.83$\pm$0.74&24.57$\pm$0.52\\
  2& 41.5857& 2 46 20.57& $-$30.2756& $-$30 16 32.35& 22.57$\pm$0.59& 22.14$\pm$0.38& 22.21$\pm$0.25\\
  3& 41.5857& 2 46 20.57& $-$30.2756& $-$30 16 32.24& 23.35$\pm$0.90& 22.95$\pm$0.68& 22.71$\pm$0.39\\
  4& 41.5859& 2 46 20.61& $-$30.2754& $-$30 16 31.66& 22.68$\pm$0.63& 23.11$\pm$0.62& 25.03$\pm$1.26\\
$\cdots$ & $\cdots$ & $\cdots$ & $\cdots$ & $\cdots$ & $\cdots$ & $\cdots$ & $\cdots$\\
\hline
\end{tabular}
}
\end{minipage}
\end{center}
\end{table*}

\begin{table}
\caption{Derived UGC 3789 ring star cluster properties}
\begin{center}
\begin{tabular}{cccccccccc}
\hline
\#&\multicolumn{3}{c}{log($t$ $\rm{yr}^{-1}$)}& &\multicolumn{3}{c}{log($M_{\rm cl} / M_{\odot}$)} & &$E(B-V)$\\
\cline{2-4} \cline{6-8}
\colhead{}&\colhead{Best}&\colhead{Min.}&\colhead{Max.}&\colhead{}&\colhead{Best}&\colhead{Min.}&\colhead{Max.}&\colhead{}&\colhead{(mag)}\\
\hline
0& 6.60 &6.60& 10.18&& 5.09& 4.56& 8.68&& 0.80\\
1& 8.46 &6.60& 10.18&& 5.78& 4.14& 7.01&& 0.20\\
2& 6.60& 6.60&  9.82&& 6.07& 5.18& 7.65&& 0.80\\
3& 6.90& 6.60& 10.07&& 5.46& 5.08& 7.59&& 0.50\\
4& 6.60& 6.60& 10.18&& 4.47& 4.01& 8.06&& 0.20\\
$\cdots$ & $\cdots$ & $\cdots$ & $\cdots$ && $\cdots$ & $\cdots$ & $\cdots$ & & $\cdots$ \\
\hline
\end{tabular}
\end{center}
\end{table}

\end{document}